%% file: TIFS_PDNPulse.tex
\documentclass[journal]{IEEEtran}
\IEEEoverridecommandlockouts
\usepackage[subtle]{savetrees}

\usepackage{tikz}
\usepackage{amsmath}

\usepackage{cite}
\usepackage{soul}
\usepackage{color}
\usepackage{fancyhdr}
\usepackage{marvosym}
\usepackage{multirow}
\usepackage{multicol}
\usepackage{booktabs}
\usepackage{amsfonts}
\usepackage[english]{babel}
\usepackage{subcaption}
\usepackage{parcolumns}
\usepackage{threeparttable}
\usepackage{duckuments}
\usepackage{ mathrsfs }
\usepackage{amsmath}
\usepackage{stmaryrd}
\usepackage{algorithm}
\usepackage{algpseudocode}
\usepackage{tabularx}
\usepackage{rotating}
\usepackage{tablefootnote}
\usepackage{xcolor,colortbl}

\usepackage{caption}
\usepackage{tikz}
\usepackage{clipboard}
\newcolumntype{C}[1]{>{\centering\arraybackslash}m{#1}}

\graphicspath{{figs/}}
\usepackage{pifont}
\usepackage{xurl}
\usepackage{hyperref}
\hypersetup{breaklinks=true}
\usepackage[draft]{fixme}
\fxsetup{layout=inline,theme=color}
\FXRegisterAuthor{ld}{ald}{\color{blue}ld}

\usepackage{listings}
\usepackage{ulem}
\usepackage{booktabs}
\usepackage{mathrsfs}
\usepackage{enumitem}

\usepackage{algorithm}

\usepackage{adjustbox}
\usepackage{array}
 
\usepackage{amsfonts}
\newcommand{\useregistered}{\iftrue}
\useregistered
\else
\fi

\newif\ifrev
\revtrue
\ifrev
  \newcommand{\huifeng}[1]{{\color{cyan} [Huifeng: #1]}}
  \newcommand{\yier}[1]{{\color{purple} [Yier: #1]}}
  \newcommand{\xiaolong}[1]{{\color{cyan} [Xiaolong: #1]}}
  \newcommand{\xuan}[1]{{\color{red} [Xuan: #1]}}
  \newcommand{\dean}[1]{{\color{green} [Dean: #1]}}
  \definecolor{persiangreen}{rgb}{0.0, 0.65, 0.58}
  \newcommand{\haoqi}[1]{{\color{persiangreen} [Haoqi: #1]}} 
\else
  \newcommand{\huifeng}[1]{}
  \newcommand{\yier}[1]{}
  \newcommand{\xiaolong}[1]{}
  \newcommand{\xuan}[1]{}
  \newcommand{\dean}[1]{}
  \newcommand{\haoqi}[1]{}
\fi

\DeclareCaptionFormat{listing}{\rule{\columnwidth}{0.1pt}\vskip1pt#1#2#3}
\usepackage{tikz}
\usepackage{pgfplots}
\usetikzlibrary{shapes,arrows,patterns}
\usetikzlibrary{pgfplots.statistics}

\newcommand*\circled[1]{\tikz[baseline=(char.base)]{
            \node[shape=circle,draw,inner sep=0.5pt] (char) {#1};}}

\hypersetup{hidelinks}
\usepackage{xspace}
\newcommand{\DesName}{PDNPulse\xspace}
\newcommand{\PDNPulse}{{PDNPulse}} %% The name of the framework

\definecolor{CommentGreen}{rgb}{0,0.6,0}
\definecolor{numbering}{gray}{0.5}
\definecolor{keywordcolor}{rgb}{0.63,0,0.42}

\lstdefinestyle{flang} {
    basicstyle=\ttfamily\footnotesize,
    numbers=none,
    fontadjust=true,
    basewidth=0.5em,
    keywordstyle={\bfseries\color{magenta}},
    commentstyle=\color{CommentGreen},
    stringstyle=\color{blue},
    tabsize=4,
    morekeywords={[1] Taint, L, H, Par, L1, L2, Domain
    }
}

\lstdefinestyle{ccode} {
    basicstyle=\ttfamily\footnotesize,
    language=C,
    numbers=none,
    frame = single,
    fontadjust=true,
    basewidth=0.5em,
    keywordstyle={\bfseries\color{magenta}},
    commentstyle=\color{CommentGreen},
    stringstyle=\color{blue},
    tabsize=4,
}

\pgfplotsset{compat=1.17}
\begin{document}
%-------------------------------------------------------------------------------

%don't want date printed
\date{}

\title{\Large \bf PDNPulse: Sensing PCB Anomaly with the Intrinsic Power Delivery Network}
% \title{\Large \bf PDNPulse: As Sensitive As Power Delivery Network for Board Anomaly Detection}

%for single author (just remove % characters)
\author{
{\rm Huifeng Zhu}\\
Washington University in St. Louis
\and
{\rm Haoqi Shan}\\
University of Florida
\and
{\rm Dean Sullivan}\\
University of New Hampshire
\and
{\rm Xiaolong Guo}\\
Kansas State University
\and
{\rm Yier Jin}\\
University of Florida
\and
{\rm Xuan Zhang}\\
Washington University in St. Louis
} % end author

\author{\IEEEauthorblockN{Huifeng Zhu\IEEEauthorrefmark{1},
Haoqi Shan\IEEEauthorrefmark{2}, 
Dean Sullivan\IEEEauthorrefmark{3},
Xiaolong Guo\IEEEauthorrefmark{4}, 
Yier Jin\IEEEauthorrefmark{2},
Xuan Zhang\IEEEauthorrefmark{1}}

\IEEEauthorblockA{\IEEEauthorrefmark{1}Washington University in St. Louis,
\IEEEauthorrefmark{2}University of Florida,
\IEEEauthorrefmark{3}University of New Hampshire,
\IEEEauthorrefmark{4}Kansas State University}}

\maketitle
\thispagestyle{plain}
\pagestyle{plain}

\begin{abstract}
%-------------------------------------------------------------------------------
The ubiquitous presence of printed circuit boards (PCBs) in modern electronic systems and embedded devices makes their integrity a top security concern.
To take advantage of the economies of scale, today's PCB design and manufacturing are often performed by suppliers around the globe, exposing them to many security vulnerabilities along the segmented PCB supply chain.
Moreover, the increasing complexity of the PCB designs also leaves ample room for numerous sneaky board-level attacks to be implemented throughout each stage of a PCB's lifetime, threatening many electronic devices. 
In this paper, we propose \textit{\DesName}, a power delivery network (PDN) based PCB anomaly detection framework that can identify a wide spectrum of board-level malicious modifications. \DesName{} leverages the fact that the PDN's characteristics are inevitably affected by modifications to the PCB. 
{\color{black}By detecting changes to the PDN impedance profile against the golden model and using the Frechet distance-based anomaly detection algorithms, \DesName{} can robustly and successfully discern malicious modifications across the system.} 
Using \DesName{}, we conduct extensive experiments on seven commercial-off-the-shelf PCBs, covering different design scales, different threat models, and seven different anomaly types. The results confirm that PDNPulse creates an effective security asymmetry between attack and defense.
\end{abstract}

%-------------------------------------------------------------------------------
\section{Introduction}\label{sec:intro}
\input{doc/intro}

\section{Background} \label{sec:background}
\input{doc/background}

\input{doc/threat_model}

\section{Proposed Detection Methodology}
\input{doc/framework}

\section{PDNPulse Analysis and Results}
\label{sec:results}
\input{doc/analysis}

\section{Defending against Adaptive Attackers}

\input{doc/discussion}

\section{Discussion and Future Work}
\input{doc/future_work}

\section{Related Work}\label{sec:relatedwork}

\input{doc/related_works}

\section{Conclusion}\label{sec:conclusion}
\input{doc/conclusion}

%-------------------------------------------------------------------------------
\section*{Acknowledgments}
%-------------------------------------------------------------------------------

Portions of this work were funded by DARPA and NSF CNS \#1739643. The views expressed in the paper are the opinions of the authors and do not represent official positions of DARPA, NSF, nor the US Government. 

% %-------------------------------------------------------------------------------
% \section*{Availability}
% %-------------------------------------------------------------------------------

% USENIX program committees give extra points to submissions that are
% backed by artifacts that are publicly available. If you made your code
% or data available, it's worth mentioning this fact in a dedicated
% section.

%-------------------------------------------------------------------------------
% \clearpage
\normalem
\Urlmuskip=0mu plus 1mu\relax
\bibliographystyle{plain}
\bibliography{./mybib}

% % \clearpage
% \section*{Appendices}
% % \addcontentsline{toc}{section}{Appendices}
% \renewcommand{\thesubsection}{\Alph{subsection}}
% \setcounter{equation}{0}
% \renewcommand{\theequation}{\thesubsection-\arabic{equation}}
% \setcounter{figure}{0}
% \renewcommand\thefigure{\thesubsection-\arabic{figure}}
% \setcounter{table}{0}
% \renewcommand\thetable{\thesubsection-\arabic{table}}  
% \input{doc/appendix}

%%%%%%%%%%%%%%%%%%%%%%%%%%%%%%%%%%%%%%%%%%%%%%%%%%%%%%%%%%%%%%%%%%%%%%%%%%%%%%%%
\end{document}

%% file: doc/intro.tex
{Modern consumer electronics brands depend on a continually growing global supply chain. The number of suppliers for a large smartphone manufacturer such as Samsung reaches into the several thousand and stretches across roughly 70 countries at more than 200 different locations~\cite{samsung2019supply}. And while it behooves each company to enlist trustworthy vendors, the sheer number of them and scope of global operations precludes thoroughly vetting everyone involved. Trustworthy vendors are paramount for the development of reliable and robust products. Nowhere is this more important than in critical infrastructures such as power grids or water treatment facilities, in privacy compliance relating to healthcare and financial institutions, or in national defense. 

Issues in the supply chain are inherently pervasive due to their scope and come with profound consequences. Direct losses and risk from counterfeiting in the global supply chain are estimated to cost billions of dollars annually~\cite{pecht2006bogus}. The United States alone, which outspends the next 11 richest countries combined on defense, allocates roughly 40\% of its military budget on electronics~\cite{dod2020budget}. A fact that has prompted recent legislation requiring both the Pentagon and Department of Defense to take steps to guarantee the security of its supply chain~\cite{smith_2021}. The issue extends beyond even reliability, with recent news of maliciously implanted microchips in Supermicro server motherboard that allowed the successful infiltration of nearly 30 companies~\cite{robertson2018big}.

Despite their ubiquity, detecting both counterfeiting and PCB Trojans remains an open problem. This is, in part, due to the complexity of the global supply chain market itself. Accounting for every potential point-of-failure or susceptibility is impractical. Efforts have been made in academia to classify state-of-the-art threats and defenses~\cite{harrison2021malicious}. Still, the work is perennially ongoing and largely unsystematic because of different attack vectors that prevent generalization. 
Although prior research has been presented on detecting counterfeits, and several robust solutions exist that focus on anomaly detection~\cite{he2017hardware} or authentication~\cite{paul2021silverin}, unfortunately, these solutions are often inherently incapable of detecting PCB Trojans. For example, embedded signature-based authentication that uses the delay of the JTAG scan chain across the PCB~\cite{paul2021silverin} can only protect complex logic ICs with the JTAG feature, and is insensitive to PCB Trojans that avoid impacting the JTAG scan chain. 
{\color{black} Several previous works~\cite{awal2022nearfield,staat2022anti}   propose using changes in radio wave propagation within an enclosed system to detect PCB in-field tampering events. However, these methods are not suitable for detecting supply chain attacks, such as Trojans or counterfeits, as they require a metal casing around the system for detection.}

To address threats in the PCB supply chain, this paper presents \textit{\DesName{}}, a novel board-level anomaly detection
framework that can identify PCB hardware Trojans, on-board counterfeit chips/components, and counterfeit PCBs. \DesName{} leverages the inherent sensitivity of the on-board power delivery network (PDN) to assure that a PCB is free from anomalies by comparing its PDN with the one of a genuine PCB. The framework relies fundamentally on the uniqueness of PDN characteristics and profiling PDN in the frequency domain.
\DesName{} can monitor a range of subtle malicious changes in the PDN impedance profile, making it ideal for accurately identifying board-level anomalies at multiple stages in the supply chain and across different systems. This capability rests on the fact that the PDN is interconnected with all subsystems on the board to provide power throughout its lifetime.  In our analysis, we have found that PCB anomalies tend to inevitably affect the PDN and are therefore detectable by \DesName{}.

A PDN in a complex PCB design often consists of subnets from several different voltage domains, with voltage regulator modules (VRMs) in each domain to supply the stable voltages, power traces from the VRMs to connect the chip pins, on-chip power grids to distribute power locally on the die, and decoupling capacitors to mitigate the voltage fluctuations at various PDN stages~\cite{bogatin2010signal}.
All PDN components are connected across multiple levels (e.g., chip, package, board) of the system and form a tree structure to create multiple voltage domains, each with its own VRMs to drive the local supply voltages~\cite{zhu2020powerscout}.
In modern electronic systems, the PDN requires low impedance to provide an adequate supply noise margin, a requirement that makes it sensitive to minute modifications. 
Hence, even minuscule changes to the PCB can affect the PDN characteristics that are detectable by accurate PDN measurements. 
On the other hand, the PDN is robust to variations in the PCB manufacturing process, which are distinguishable from malicious changes when profiled in the frequency domain.
%, thereby ensuring the robustness of the \textit{PDNPulse} framework.
%On the other hand, as will be shown later, compared with the changes due to the PCB design modification, the process variation of PCB negligibly affects PDN's characteristics, making PDNPulse a robust framework.

To the best of our knowledge, \PDNPulse{} is the first general board-level anomaly detection method that can be used for monitoring full-system, cross-layer behavior. {Several recent approaches~\cite{nishizawa2018capacitance, wang2019system} have proposed PDN impedance-based detection, but do not achieve general, full-system, or cross-layer anomaly detection. Nishizawa et al.,~\cite{nishizawa2018capacitance} models the PDN as a resistor and capacitor in parallel, and measure the PDN capacitance by injecting a single-frequency sine wave and measuring the corresponding amplitude of the current. They demonstrate detection of an anomalous capacitor as low as $0.1\mu F$. Wang et al.,~\cite{wang2019system} indirectly measure the PDN impedance at resonant frequency to detect counterfeit PCBs by proposing a ring oscillator (RO) array embedded in integrated circuit (IC). They leverage the fact that when the IC is clocked at the resonant frequency of the PCB's PDN (e.g., red dot in Fig.~\ref{fig:voltage_domain}(c)), the observed supply voltage fluctuations are mainly affected by the PCB PDN impedance. This is measurable as changes in the oscillation frequency of the embedded RO and allows anomaly detection by comparison with a trusted database. This solution further allows authentication by clocking the IC at non-resonant frequencies so that the embedded RO acts as a physically unclonable function (PUF).

\DesName{}, on the other hand, analyzes the PDN using a complex model that includes the parasitics of each on-board component (see Fig.~\ref{fig:pds_methodology}(c)) using the complete PDN impedance profile in the frequency domain. 
As a result, we demonstrate significantly improved detection sensitivity of anomalously inserted capacitance as low as $1.8fF$.\footnote{The value is an order-of-magnitude smaller than the parasitic of the 3pin SOT-23 package, one of the smallest chip package footprints (2.6mm$\times$2.9mm).}
Both~\cite{nishizawa2018capacitance,wang2019system} are currently only capable of detecting anomalies in one voltage domain using one (or two) frequency points. 
Anomalies in other voltage domains or placed far from the measured ports in those defenses can possibly evade detection due to isolation effects of the VRMs and decoupling capacitors, or motivated attackers that shift the affected frequency band of their malicious insertion to bypass detection.
To solve this challenge, we extend the traditional PDN analysis~\cite{gupta2007understanding} by using multi-domain multi-port detection and measuring both self- and transfer PDN impedance.
What is more, based on the designed probe, \DesName{} does not require any embedded, or otherwise, hardware modifications. 
This feature allows flexibility of detection within varying phases of the supply chain, in the field, and on legacy systems without an existing defense.  
In addition, detection in \cite{nishizawa2018capacitance,wang2019system} fails to recognize that in frequency-domain
PDN analysis, malicious modifications with minimal parasitics are still observable as shifted PDN profiles with discernible
magnitudes~\cite{kim2010chip} at high frequencies. This property facilitates
PDNPulse’s detection sensitivity and allows us to employ a pattern-based method using Frechet distance, decreasing the possibility of evasion and increasing the robustness to PCB process variations. } 

In general, we demonstrate that \DesName{} provides
broad assurance for commercial-off-the-shelf (COTS) electronics across all design layers in the supply chain. %, while at the same time being light-weight and compatible with legacy systems by reusing the intrinsic feedback/sensing capabilities in the PDN. 
%Our results show that \textit{\PDNPulse} can successfully detect a wide range of sophisticated board-level attacks encompassing hardware Trojans, counterfeiting, and recycling. 
% Our analysis comprehensively demonstrates rapid identification and detection of the majority of board-level attacks.
The contributions of this work are summarized as follows:
\begin{itemize}%[leftmargin=*]
    
    \item We coalesce different board-level attack effects using PDN impedance profiling and multi-port, multi-domain PDN measurement methodologies for systematic detection.
    
    \item We propose \PDNPulse, the first general board-level attack detection framework used for monitoring full-system, cross-layer behavior. We present the workflow of \PDNPulse, providing comprehensive setup and procedural guidance. We also design a custom probe for \PDNPulse{} that achieves good trade-offs among accuracy, error, and ease-of-use. 
    
     \item We develop a modified Frechet distance to evaluate the minute differences between the PDNs on two PCBs, which also serves as the security metric. Based on our approach, we develop robust algorithms for both anomaly detection and board classification.
    
    \item We present extensive experimental results and analysis of \PDNPulse{} on a wide-range of custom and COTS PCBs. In so doing, we cover different design scales and attack types that demonstrably validate the sensitivity and robustness of \PDNPulse{} for the majority of board-level attacks.
    
\end{itemize}

%% file: doc/background.tex
\begin{figure}[t]
\centering
\includegraphics[width=3.45in]{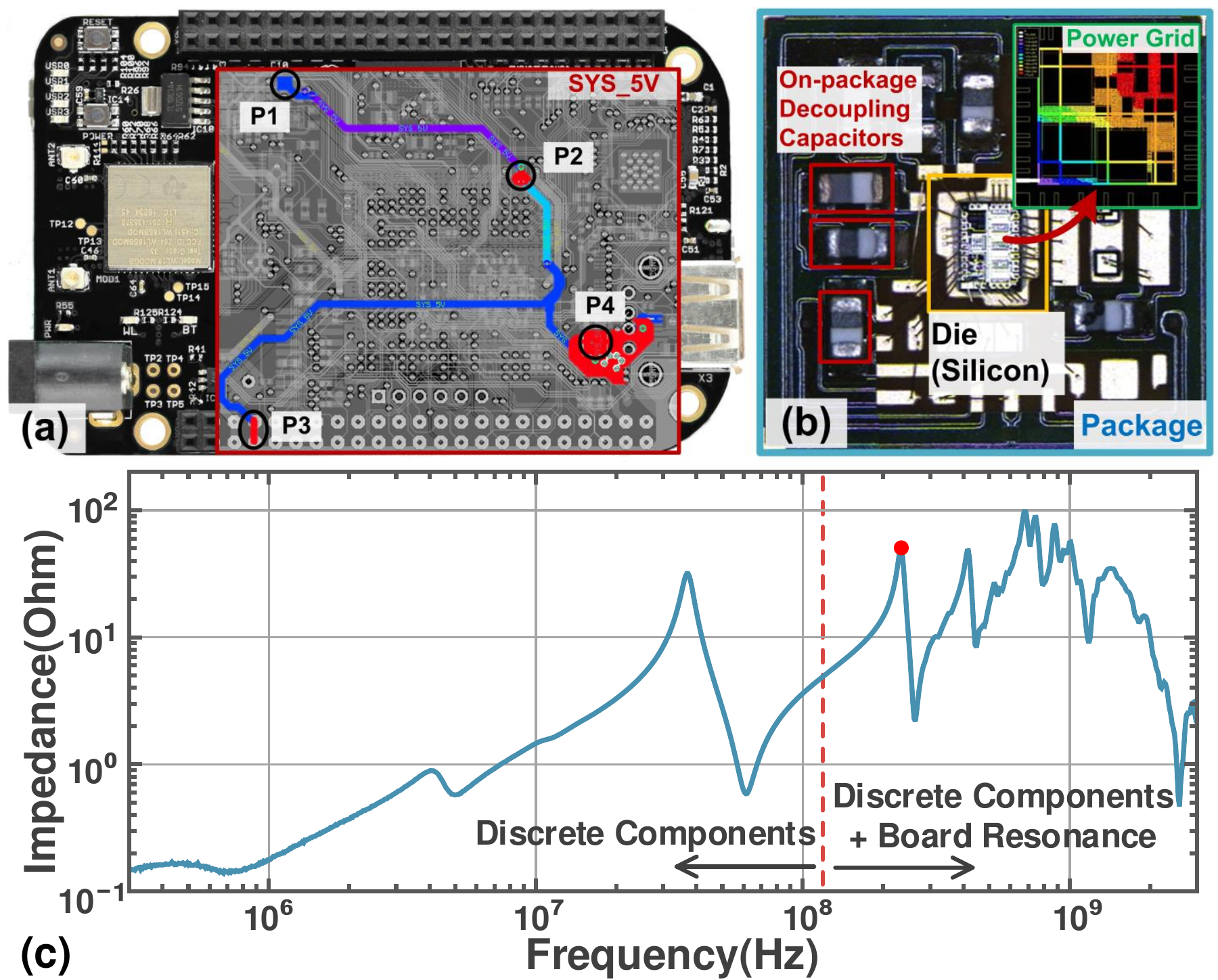}
\caption{(a) One voltage domain (\texttt{SYS\_5V}) of an example system~\cite{beaglebone_wireless}, with the power supply net highlighted. P1-P4 specify accessible probe points for measuring this  domain's PDN. (b) A chip-level PDN, with in-package decoupling capacitors and die-level power grid highlighted. (c) An example PDN impedance profile (magnitude of $Z_{11}$) of a custom experimental board.}
\label{fig:voltage_domain}
% \vspace{-0.1in}
\end{figure}

\subsection{Power Delivery Network (PDN)} \label{sec:background_PDN}
The PDN must provide a stable supply
voltage and sufficient power to other on-board modules.
In a complex PCB design, chips/components have different power distribution requirements for reliable operation, such as supply voltage levels, maximum load currents, and voltage noise margins.
Thus, the PDN is composed of VRMs that form a tree structure to create multiple voltage domains.
Fig.~\ref{fig:voltage_domain}(a) shows one of 17 voltage domains of a BeagleBone single board computer~\cite{beaglebone_wireless}, highlighting its power supply net.

Different voltage domains have different power supply specifications and cover different chips/components on the PCB.
{ Each voltage domain has its own VRM to convert the power supply from the upper-node voltage domain and drive the local supply voltage to the chips.
The VRM also isolates the two voltage domains since one of its primary functions is to prevent the voltage fluctuations of one domain from propagating to the other domain. When powered off, the VRM is an open circuit, thereby disconnecting the two domains.
}
Between the hierarchical VRMs and chips is the board-level passive distribution network, containing PCB power wire lines, power planes, and on-board discrete decoupling capacitors.
% Specifically, the power planes are flat planes of copper that occupy the empty area of each layer of the PCB beside the signal wires and are connected to the PDN.
Given a voltage domain, probes can be attached to the accessible points (e.g., P1-P4 of Fig.~\ref{fig:voltage_domain}(a)) of the power supply net to detect disturbances. 
Measurements taken at these points can isolate the impedance profile of, for example, \texttt{SYS\_5V}, from the 17 possible voltage domains. 
At the chip level, Fig.~\ref{fig:voltage_domain}(b) shows both power grids to distribute power locally on the die and decoupling capacitors at the die or package level.
All PDN components are connected across multiple levels (i.e., die, package, and board) of the system and form an infrastructure that can sense disturbance within the system.

\subsection{PDN Impedance Profile}
%\textit{PDNPulse} relies on detecting minute changes of the PDN impedance profile, which is the impedance of PDN seen in the frequency domain.
The impedance profile (also known as \textbf{$Z$-parameters}) of a PDN in the frequency domain is widely used to evaluate its performance, { which are represented as a symmetric matrix.
% \footnote{See Appx.~\ref{sec:impedance_profile} for the definition of $Z$-parameters.}:
\begin{equation}
\footnotesize
Z_{PDN}(f)=\begin{bmatrix}
    Z_{11} & Z_{12} & \cdots & Z_{1n}\\
    Z_{21} & Z_{22} & \cdots & Z_{2n}\\
    \vdots & \vdots & \ddots & \vdots\\
    Z_{n1} & Z_{n2} & \cdots & Z_{nn}
    \end{bmatrix}
    \label{eq:z-matrix}
\end{equation}
\noindent where $n$ is the number of measured ports, diagonal elements $Z_{xx}$ are the self-impedance seen from each measurement port, and non-diagonal elements $Z_{xy}$ are the transfer impedance between two ports.
For traditional PDN analysis, only self-impedance is of interest since it can represent the quality of power supplied to a chip.}
{While for PDN-based anomaly detection, we should also focus on the transfer impedance since on-board capacitors behave like barriers, separating one voltage domain into multiple subdomains. Self-impedance can precisely characterize the PDN in one subdomain, while transfer impedance can sense across multiple subdomains at the cost of higher noise. By combining self-impedance and transfer impedance, we can increase the overall detection accuracy and sensitivity.}

Fig.~\ref{fig:voltage_domain} (c) is an example profile of PDN self-impedance ($Z_{11}$).
Both self- and transfer impedance profiles can be roughly divided into two parts:
1) the low-frequency part, which is due to the electrical characteristics
of the PDN circuit (specifically, discrete components), and 2) the high-frequency part, which is mainly due to the
electromagnetic resonance formed by the PCB cavity between the power planes (i.e., board resonance). 
Both parts of the profile help reveal the effects introduced by PCB anomalies concerning circuit-level changes and board resonance changes, respectively.
Combining both the circuit level and board resonance information, the impedance profile captures minute changes in the PCB design, even if those changes do not directly impact the operability of the PDN circuit itself. 
Throughout this paper, both low-frequency and high-frequency information are used together to detect PCB modifications.

%% file: doc/threat_model.tex
\section{Threat Model}\label{sec:threat_model}
\noindent\textbf{Attack Surface.} Attackers can perform physical modifications during any stage of the PCB's life cycle, such as design, fabrication, integration, distribution, and repair.
They have full access to the PCBs and their design details, such as the schematic, layout, and bill-of-material (BoM).
The intermediate parties or legitimate end-users can use \DesName{} to detect anomalies on populated (i.e., with all components assembled) PCBs and to verify the trustworthiness along the supply chain.

% They can implant anomalies by inserting hardware Trojans or performing counterfeit (including low-quality and recycled) replacements. 

\smallskip
\noindent\textbf{Attackers' Motivation.} Attackers are dishonest opportunists driven by financial or security incentives. Their goal is to gain either profit or valuable information. 
Practical PCB threats need to be stealthy (i.e., no blatant violation of design rules or functional failure) and meaningful (i.e., no frivolous modifications without security or financial gains). Thus, attacks that uncontrollably compromise the basic functionality (e.g., short circuits) or have insignificant security impacts (e.g., moving a single via) are out of the scope of our work. 
{Attackers that can undo the changes (e.g., remove the malicious plug-in before \DesName{}'s detection) are also not in scope.}
Further, implementing Trojans by exclusively modifying a chips' internal structure (i.e., chip-level Trojans inserted by attackers in chip supply chains) are not considered. 

\smallskip
\noindent\textbf{Attack Vectors.} 
Attackers can maliciously yet meaningfully add, remove, alter, and replace arbitrary electrical components of the PCB.
Specifically, attackers can implant anomalies by inserting Trojan circuits or performing counterfeit (including low-quality and recycled) replacements. 
We show that \DesName{} can effectively detect the majority of practical board-level attacks, as summarized below:

\begin{itemize}[leftmargin=*,topsep=0px,partopsep=0px]
    \item \textbf{PCB Trojans (Sec.\ref{sec:trojan_detection_results}).} {Trojan circuits create a backdoor for attackers and can be utilized to launch attacks compromising security assurance.} Known practical PCB Trojans~\cite{zhu2021pcbench,harrison2021malicious} fall into two main categories:
    
    \begin{description}
    \item\textbf{\textit{Triggerable Trojans (Sec.\ref{sec:pdnpulse1v3} and \ref{sec:arduino_due_board}).}}
    % They represent the most common type of PCB Trojans~\cite{harrison2021malicious,mehta2020big}, which are based on small-package chips.
    At the board level, Triggerable Trojans are based on small-package chips.
    For example, chips that integrate numerous logic gates are implanted to be highly functional yet sneaky.
    To achieve advanced attacks with complex trigger patterns or payload functions, processor chips (e.g., microcontrollers (MCU)) are commonly adopted~\cite{greenberg2019tinyspy}.
    
    \item
    \textbf{\textit{Always-on Trojans (Sec.\ref{sec:pdnpulse1v3}).}} One notable attack is to steal sensitive information (e.g., secret keys) on the chips by inserting sampling resistors in the power rails that can perform side-channel analysis attacks~\cite{flynn2018power,wei2018know}.
    
    % \item
    % \textbf{\textit{Debug Ports Snooping (Sec.\ref{sec:pdnpulse1v3}).}} Adversaries can manipulate unprotected debug ports to gain absolute privilege for performing illegal activities~\cite{exploitee2017all}.
    \end{description}

\item  \textbf{Chip/Component Counterfeits (Sec.\ref{sec:counterfeit_chip}).} {The security of such components are unverified. Thus they can be leveraged by attackers to launch attacks (e.g., inject faults when running code).} In this paper, we focus on two main types of on-board counterfeits~\cite{guin2014counterfeit}:

\begin{description}
\item\textbf{\textit{Counterfeit Chips (Sec.\ref{sec:pynq}).}} We refer to chips as those with programmable functions, such as MCUs, microprocessors, and field programmable gate arrays (FPGAs).

\item
\textbf{\textit{Counterfeit Components (Sec.\ref{sec:motherboard}).}} Other chips are of this type. Examples include transistors, logic gates, and amplifiers. Passive components (e.g., resistors) are not considered since there exists no known practical demonstration of their profitability in counterfeiting.
\end{description}

\item \textbf{PCB Counterfeits (Sec.\ref{sec:counterfeit_board}).} {Such counterfeits expose systems to vulnerabilities and increased failure rates.} Practical counterfeit PCBs are usually of three types with varying degrees of stealth:

\begin{description}
\item
\textbf{\textit{Imitating  (Sec.\ref{sec:i350t4}).} }
    Attackers have complete access to the PCB design resources. However, for higher profits, they typically replace parts of the original circuit with a low-standard design.
    % Attackers have access to only part of the PCB design resources. 
    The fabricated counterfeit boards are thus different from the original PCBs, which is sometimes observable from the board layout.
    Still, they can remain undetected since the boards are usually inside the products (e.g., servers), preventing imaging inspection~\cite{comparison_intel_i350}.
\item
\textbf{\textit{Cloning (Sec.\ref{sec:arduino_uno}).}} Adversaries also have all PCB design information. They can fabricate the board with the same layout while embedding counterfeit or low-quality components~\cite{mark2017honda}. This type of counterfeit can be quite difficult to visually distinguish from a genuine board.
% \item
% \textbf{\textit{Recycling (Sec.\ref{sec:arduino_uno}).}} Used PCBs are recycled and sold as brand-new ones. This is the stealthiest type since recycled boards can have the same appearance as genuine PCBs.
\end{description}

\end{itemize}

\smallskip
\noindent\textbf{Golden Model.} 
Note that \DesName{} fundamentally identifies whether the tested board can be trusted.
In this work, genuine boards (i.e., the golden model) can be those supplied directly from the original equipment manufacturer (OEM), including the PCB itself and its on-board electronic components. 
Genuine boards can also be achieved by selecting one PCB and conducting reverse engineering.
When building the golden model, process variations due to both PCB fabrication and component variation need to be considered. Specifically, component variations can arise from either fabrication tolerances or the adoption of multiple BOMs.
If the deviation of the PDN impedance profiles exceeds the recorded tolerance for the golden model, we can reasonably regard a board under test as untrusted (i.e., malicious/counterfeit/suspicious). Our method is not designed to pin point the root causes or malicious intent of such deviations.

%% file: doc/framework.tex
In this section, we first introduce the overall workflow of the proposed \DesName{} framework. We then elaborate on how \DesName{} can help detect different board-level attacks,
% highlighting how we analyze PDN sensitivity, 
% i.e., our algorithmic approach for delineating minute changes in impedance profiles and unifying anomaly detection,
and discuss the challenges and considerations when measuring the impedance profile of a PCB with respect to attack vectors. 

\subsection{PDNPulse Framework} \label{sec:experimental_setup}

% \begin{figure}[t]
% \centering
% \includegraphics[width=2.9in]{figs/setup_probe.pdf}
% \caption{(a) The experimental setup for measuring the PDN impedance profile. (b) The customized probe for low-noise PDN impedance measurement.}
% \label{fig:setup}
% \end{figure}

\begin{figure}[t]
\centering
\includegraphics[width=3.45in]{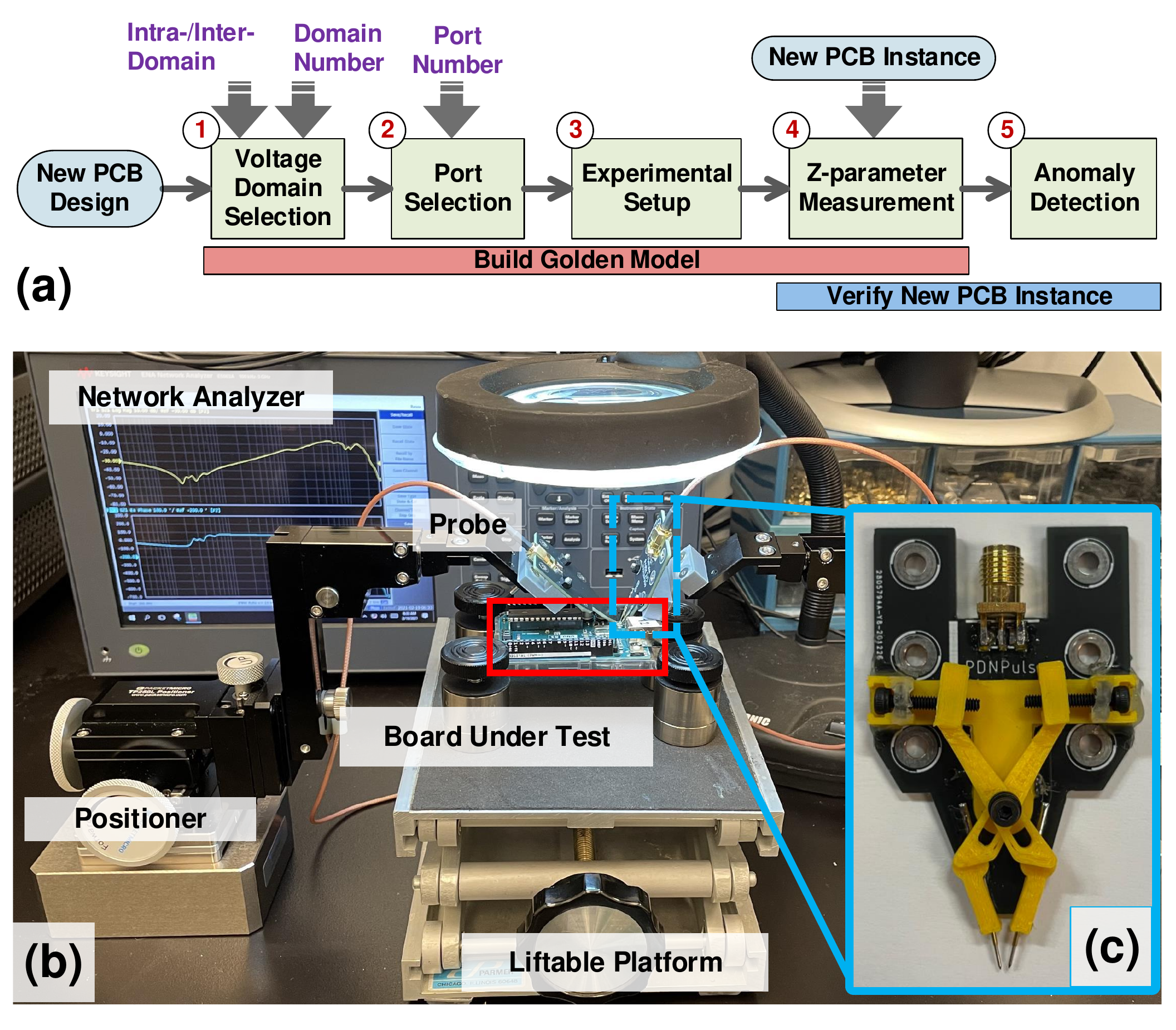}
\caption{(a) Proposed PDNPulse framework. (b) The experimental setup for measuring the PDN impedance profile. (c) The customized probe for low-noise PDN impedance measurement.}
\label{fig:framework}
% \vspace{-0.1in}
\end{figure}

The components of the \DesName{} framework are shown in Fig.~\ref{fig:framework}(a), in a step-by-step manner. {To build the golden model of a new PCB design, Steps 1$\sim$3 should be done once, followed by Step 4 to record several genuine PCB instances. To verify new PCB instances, users should then only conduct Steps 4 and 5.}

\smallskip
\noindent\textbf{\circled{1} Voltage Domain Selection.} A PCB's PDN is comprised of multiple voltage domains. Each voltage domain corresponds to a set of components connected to this domain and the PCB region of this domain. 
% The voltage domains need to be first identified. 
\DesName{} applies multi-domain detection, obtaining measurements from several voltage domains.  Selected voltage domains determine the detection coverage. Complete coverage can be achieved by measuring all voltage domains. 
% Given the multi-domain method, measurement ports can be selected to be either within the same domain (intra-domain) or from different domains (inter-domain). 
Given the multi-domain method, measurements can be either intra-domain or inter-domain. 
In intra-domain detection, to achieve low noise, each voltage domain is measured separately without including the interaction between voltage domains.
On the other hand, in inter-domain detection, multiple domains are measured together and the coupling effects between different voltage domains are analyzed. This scheme is especially useful for detecting minute changes in chip/components, because the coupling effects  are much more pronounced at the chip level than at the PCB level.
Inter-domain detection is also applicable when there are insufficient testing ports in one target voltage domain.

\smallskip
\noindent\textbf{\circled{2} Port Selection.} 
In this step, users choose the number of testing ports, whose locations are based on the detection targets. 
Note that testing more ports increases the overall anomaly detection performance, but it also increases the testing cost. 
The selected ports must be the reachable points of the power supply net (see Fig.~\ref{fig:voltage_domain}(a)).
One rule of thumb for the best performance is to avoid directly placing the probe at low-impedance nodes (e.g., next to decoupling capacitors), because the measured impedance will be dominated by this low impedance, which overshadows the PDN profile we intend to measure, causing large distortion. 
% Also note that the best result is obtained when the probe point is close to the anomaly site.
% In practice, when facing real-world attacks, users may not be able to precisely locate the anomaly, which motivates the adoption of multi-port detection in our \DesName{} framework to cover different parts of the PCBs. 

\smallskip
\noindent\textbf{\circled{3} Experimental Setup.} Once the port locations are determined, the target PCB is put on the testbed, along with positioners, a vector network analyzer (VNA), and multiple probes for measurement.
The VNA is configured with the standard 2-port shunt-through method for small impedance measurements~\cite{sandler2016extending}, followed by standard 2-port calibration to obtain high-precision measurements.
% \footnote{Although our probes are not calibrated at the tip end, the systematic error will not affect the findings from the results. This error can be further removed if calibration is conducted at the probe tip end. }.
% Details of the calibration can be found in Appx.~\ref{apx:claibration}.
% (See Fig.~\ref{fig:setup}). 
%The setup includes the setup of testbed and VNA parameters.
%The testbed used in this paper is shown in Fig.~\ref{fig:setup} and VNA setup will be introduced in Sec.\ref{sec:experimental_setup}.
%Fig.~\ref{fig:setup}(a) shows the adapted testbed throughout this paper. We utilize the Keysight E5063A VNA and take the frequency domain $300KHz\sim 3GHz$ as the band of interest. 

\smallskip
\noindent\textbf{\circled{4} $Z$-Parameter Measurement.} The $S$-parameters of the abstracted PDN are first measured by VNA, which are then converted to $Z$-parameters (i.e., impedance profiles).
One challenge here is that existing method~\cite{sandler2016extending} is can only be used in measuring self-impedance. 
To solve this challenge, we extended the previous method~\cite{sandler2016extending}. 
Take the PDN with two ports as an example, we first measure the self-impedance (i.e., $Z_{11}$ and $Z_{22}$) of each port using the classic method~\cite{sandler2016extending}.  
Then, two ports of the VNA are connected to the two ports of PDN, respectively, to measure $S_{21}$. Based on the equivalent circuit, we can calculate the transfer impedance $Z_{21}$ ($Z_{12}$)~\cite{novak2007frequency}:
\begin{equation}
\footnotesize
    Z_{21}=Z_{12}=S_{21}\frac{Z_0}{2}\frac{1+\frac{Z_{11}}{Z_0}+\frac{Z_{22}}{Z_0}+\frac{Z_{11}}{Z_0}\frac{Z_{22}}{Z_0}}{1+S_{21}\frac{Z_{11}}{2Z_0}}\label{eq:s2z_matrix}
\end{equation}
\noindent
where $Z_0$ is the reference impedance of VNA. 
By repeating such a process for every two ports, we can obtain the complete $Z$-parameters. Note that all measurements are performed off-line without powering the boards.

% \begin{figure}[h]
% \centering
% \includegraphics[width=3in]{figs/imped-matrix.pdf}
% \caption{The process of measuring the PDN Z-parameters using the 2-port shunt-through method.}
% \label{fig:processZmatrix}
% \end{figure}

\smallskip
\noindent\textbf{\circled{5} Anomaly Detection.} In this step, users apply anomaly detection algorithms (introduced in Sec.\ref{sec:algorithm}) to the generated impedance profiles to determine if any anomaly exists.
% and what type of the anomaly is detected.
% In addition, for the new PCBs with the same design, we can start with the experimental setup and $S$-parameter measurement. \yier{I am lost on this paragraph. Are we talking about S measurement or probe?}

\medskip
Fig.~\ref{fig:framework}(b) shows the experimental setup used throughout this paper. We utilize the Keysight E5063A VNA and take the $300KHz$-$3GHz$ as the band of interest, which is the typical selection for PDN analysis. 
The VNA is set to $10KHz$ maximum intermediate frequency (\texttt{IF}) bandwidth, and 
1024 points are collected for each impedance profile to obtain a proper frequency resolution. 
% In addition, using positioners is essential for measurement stability and efficiency.

While most components for our testbed were commercially available, the probe was a custom design (see Fig. \ref{fig:framework}(c)) to meet our unique requirements\footnote{We release the source files for our probe design: \url{https://github.com/xz-group/PDNPulse/tree/main/probe}.}.
% In addition to introducing minimum parasitic effects, 
% the space between the probe tips is adjustable for measuring ports at various distances.
We implemented short $50\Omega$ traces to reduce the probe's parasitic effects and match the impedance of VNA.
The traces connected two probe tips (signal and ground) with springs to a coaxial adapter, which can be further connected to the VNA. 
A mechanical probe tuner (the yellow part in Fig.~\ref{fig:framework}(c)) was also designed to precisely adjust the space between the probe tips for measuring ports at various distances. 
This design overcomes the limitation of commercial probes with fixed tip spacing. 
The distance between the tips can be adjusted from 0.05mm to 4mm by a screwdriver. 
Using standard calibration method~\cite{SOLT}, the probe can be compensated within 3GHz bandwidth, which is the setup throughout this paper.
In addition, each probe costs around \$20.
Overall, our probe is accurate, relatively inexpensive, and easy to use.

\subsection{PDN Sensitivity} 
\label{sec:pdn_sensitivity}

% Our method eventually leverages the fact that any PCB anomaly will inevitably introduce impacts on the PDN impedance profile due to the parasitic effects of any malicious modifications. 
% By measuring, analyzing, and comparing the impedance profiles of target PCBs, we can achieve the goal of PCB anomaly detection with high sensitivity.

\begin{figure}[t]
\centering
\includegraphics[width=3.45in]{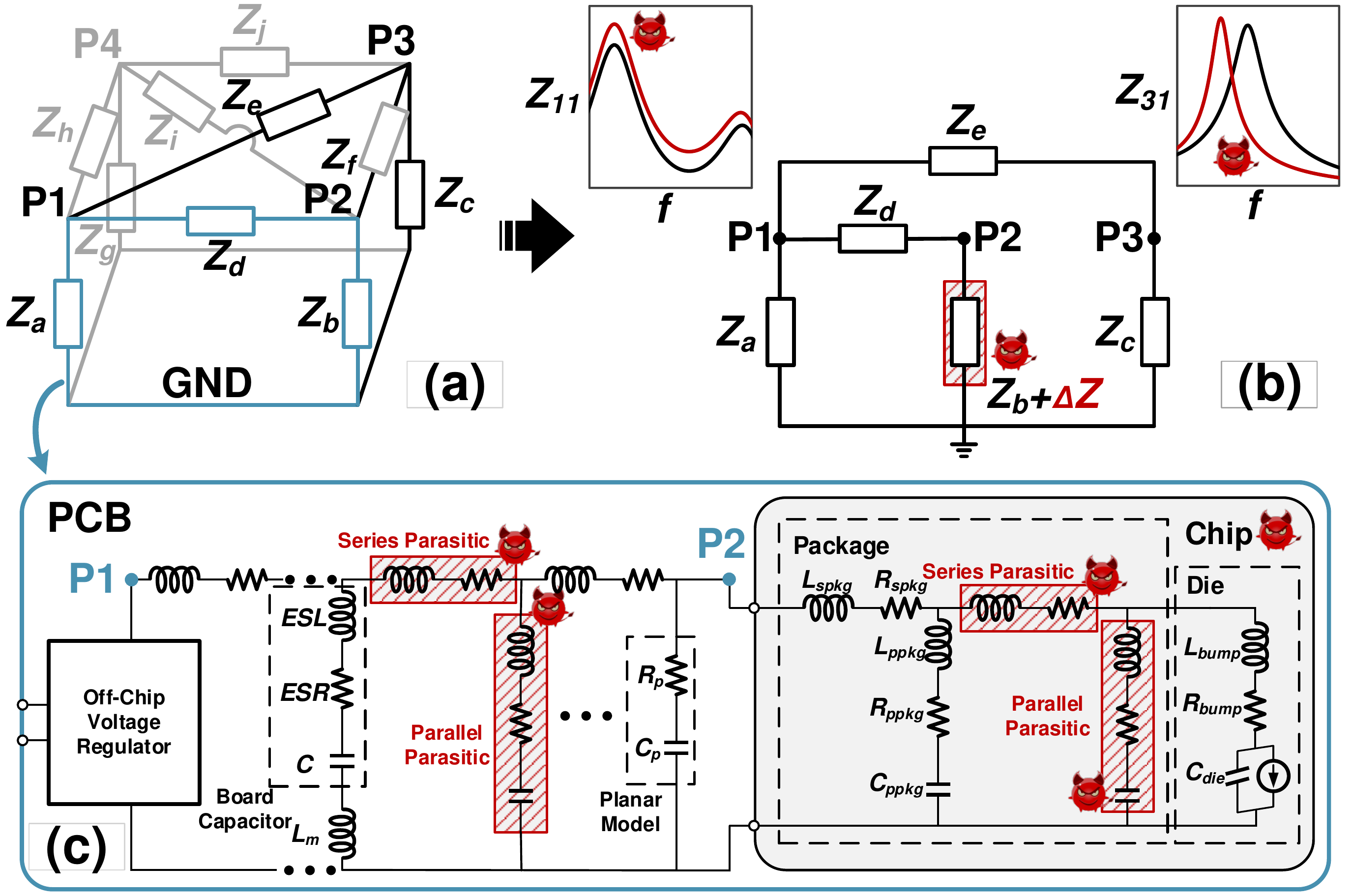}
\caption{(a) A conceptual network of a PDN and (b) its equivalent circuit showing the PDN sensitivity, where $Z_b$ is maliciously modified, denoted by $Z_b$$+$$\Delta Z$. The effect of the malicious modification on the impedance profile of $Z_{11}$ and $Z_{31}$ is shown in red. (c) Simplified circuit model of the PDN with malicious modifications outlined.}
\label{fig:pds_methodology}
% \vspace{-0.1in}
\end{figure}

Our method leverages the inherent sensitivity of the PDN for board-level anomaly detection. Fig.~\ref{fig:pds_methodology} illustrates the key idea of our method. The PDN of the PCB can be viewed as a three-dimensional impedance network, as shown in Fig.~\ref{fig:pds_methodology}(a), where the top side is the power supply and the bottom side is the ground. 
The entire PDN is then abstracted into multiple ports, e.g., the four ports (P1, P2, P3, and P4) in Fig.~\ref{fig:pds_methodology}(a). Between any two ports and between each port and the ground, there exists an impedance component, namely, $Z_a\sim Z_j$. For each impedance component, there can be a subnetwork of a series of resistors (R), inductors (L), and capacitors (C).

To show how each component can affect the impedance profile, we show a simplified example without loss of generality. That is, we ignore P4 and assume P2 has no direct connection to P3 (i.e., $Z_f$$=$$\infty$). Thus, we can achieve the equivalent circuit shown in Fig.~\ref{fig:pds_methodology}(b). According to the $Z$-parameter's definition, the diagonal and non-diagonal $Z$-parameters in Eqn.~(\ref{eq:z-matrix}) can be calculated as follows~\cite{wiki_parallel}:%\footnote{All six $Z$-parameters are listed in the Appx.~\ref{apx:full_list_z}.} 
\begin{gather}
\footnotesize
    Z_{11}=Z_a\parallel (\boldsymbol{Z_b}+Z_d)\parallel (Z_c+Z_e)\label{eq: example_z11}\\
    % Z_{22}=Z_b\parallel (Z_d+Z_a\parallel(Z_c+Z_e))\\
    % Z_{33}=Z_c\parallel (Z_e+Z_a\parallel(Z_b+Z_d))\\
    % Z_{12}=Z_{21}=\frac{Z_a\parallel(Z_c+Z_e)+Z_b+Z_d}{Z_b(Z_a\parallel(Z_c+Z_e))}\\
    \footnotesize
    Z_{13}=Z_{31}=\frac{Z_a\parallel(\boldsymbol{Z_b}+Z_d)+Z_c+Z_e}{Z_c(Z_a\parallel(\boldsymbol{Z_b}+Z_d))}\label{eq: example_z13}
    % Z_{13(31)}=[Z_a\parallel(\boldsymbol{Z_b}+Z_d)+Z_c+Z_e ]/[{Z_c(Z_a\parallel(\boldsymbol{Z_b}+Z_d))}]\label{eq: example_z13}
    % Z_{23}=Z_{32}=\frac{Z_c(Z_a\parallel(Z_c+Z_e))(Z_b\parallel (Z_d+Z_a\parallel(Z_c+Z_e)))}{(Z_a\parallel(Z_c+Z_e)+Z_d)(Z_c+Z_e)}
\end{gather}
\noindent %\yier{Why $Z_b$?}
where $Z_b$ is highlighted and works as an example. %\yier{you only show two Z parameters for simplicity. Is this simplicity due to the ignoring P4 or just for space consideration?}\huifeng{For space consideration.}\yier{Then, put them in the Appendix.}
As shown in Fig.~\ref{fig:pds_methodology}(b), if $Z_b$ changes due to the malicious modification, each $Z$-parameter element of the network will change in a complex way, as modeled by Eqn.~(\ref{eq: example_z11}) and (\ref{eq: example_z13}). The pattern and amplitude of the impedance profile ($Z$-parameter) changes depend on the topology of the PDN and the values of each impedance component.

The example in Fig.~\ref{fig:pds_methodology}(b) intuitively illustrates the PCB PDN sensitivity. Although the real PDN is much more complex than the above example, Fig.~\ref{fig:pds_methodology}(c) shows a simplified PDN schematic, which is part of the entire PCB PDN. In this figure, all electronic components, such as discrete capacitors, power planes, and on-board chips, can be modeled by RLC components. Together, they form a complex network where each of the PDN $Z$-parameters can be represented as below:
\begin{equation}
\footnotesize
    Z_{xy}=f(\{R_1,...,R_l\},\{L_1,...,L_m\},\{C_1,...,C_n\}) \label{eq:zxy_rlc}
\end{equation}
\noindent
where $R_i$, $L_j$, and $C_k$ include the parasitic effects of all PDN
components. Anomalies at different levels will inevitably introduce changes to the original PDN due to the parasitic effects of the modifications, which can also be modeled as exogenous RLC components, as shown in the red-marked regions in Fig.~\ref{fig:pds_methodology}(c). The values of $R$, $L$, and $C$ in Eqn.~(\ref{eq:zxy_rlc}) will then deviate from the original values. 
%As a result, the PDN impedance profile will change accordingly.

\subsection{Frechet Distance-based Anomaly Detection Algorithms} \label{sec:algorithm}

Note that changes in the impedance profile are mainly due to the parasitic effects of a PCB anomaly, causing a shift of the impedance profile, as shown in red in Fig.~\ref{fig:pds_methodology}(b). Besides, different anomalies can affect the impedance profile at different frequency bands. To facilitate unified
anomaly detection based on the PDN, we focus on the impedance profile pattern instead of merely comparing the impedance amplitudes at a specific frequency. 
% A reliable way to detect the anomaly is mapping to the circuit model from the impedance profile, i.e., deriving the R, L, and C values in Eqn.~(\ref{eq:zxy_rlc}) according to the impedance profile. Then the anomaly
% detection is based on examining the changes of RLC values.
% However, the mapping process is not robust to the measurement noise. 
% Besides, it is challenging to make sure the mapped circuit models from different $Z$-parameters of the same PDN are consistent with each other, thus enabling multi-port detection. which is out of the scope of this work. \xuan{this discussion about the method we are not using is unclear. why do we care?}
% Therefore, we adapt the Frechet
% distance~\cite{eiter1994computing} as the impedance similarity score to evaluate the similarity between the patterns of two impedance profiles. The Frechet distance ($FD$) between two curves is defined as: 
Therefore, we adopt the Frechet
distance~\cite{eiter1994computing}, which measures the similarity between two curves, as the security metric to evaluate the difference between the impedance profiles and to quantify the uniqueness and stability of \DesName{}. The Frechet distance ($FD$) is defined as: 
\begin{equation}
\footnotesize
    FD(A,B)=\inf_{\alpha,\beta} \max_{t\in[0,1]}\Bigl\{d\bigl(A(\alpha(t)),B(\beta(t))\bigr)\Bigr\}\label{eq:frechet}
\end{equation}
\noindent
where $A$ and $B$ are the two curves, $\alpha(t)$ and $\beta(t)$ are arbitrary continuous non-decreasing functions, and $d$ is the Euclidean distance between two points of the two curves.
When $t=0$ and $1$, $\alpha(t)$ and $\beta(t)$ are mapped to the endpoints of the curves.
Frechet distance takes into account the location and ordering of the points along the curves when measuring similarity.
%A popular illustration of the Frechet distance is assuming the man is walking on one curve and the dog on another curve. Both can adjust their speeds but are not allowed to move backwards. The Frechet distance of the two curves is then the minimum length of leash necessary to connect the man and the dog~\cite{alt1995computing}. 
It has been used to distinguish electromechanical impedance curves in the materials science field~\cite{soman2020development,Jekel2019}.

For multi-port based PDN measurement, we propose to calculate the $FD$ between two boards (noted as $FD'$) as the norm of the $FD$s for each port:
\begin{equation}
\footnotesize
FD'(B_1,B_2)=\frac{||\{FD(\log_{10}(Z_{xy}^{B_1}),\log_{10}(Z_{xy}^{B_2}))\}_{x\in[1,n],y\in[1,n]}||_{m}}{(n^2+n)/2} \label{eq:fd_multi_ports}
\end{equation}
where $B_1$ and $B_2$ are two boards, $n$ is the number of ports, and $m$ is the order of the norm. The factor $(n^2+n)/2$ is the number of $Z$-parameters for an $n$ port network, which is used to normalize the $FD'$. The $FD'$ is computed with the $Z$-parameters in $log_{10}$ scale to avoid the high-frequency part dominating $FD$. Commonly selected norms include the L1 norm, L2 norm, and uniform norm. 

The proposed $FD'$ can be viewed as an analog domain alternative of the Hamming Distance (HD) that is widely used for system identification~\cite{bhunia2018hardware}.
% For PUFs, there are two important security metrics: stability (i.e., intra-HD) and uniqueness (i.e., inter-HD). 
% The former represents the ability to generate the same output for the same input under different environmental conditions, and the latter is, the distinguishability of a PUF with respect to other instances. 
% However, different from PUFs, for \DesName{}, the stability (i.e., intra-$FD'$) is calculated as the $FD'$ among the Reference (also noted as Genuine) boards to capture the PCB process variation. 
% The uniqueness (i.e., inter-$FD'$) is calculated as the $FD'$ between Reference and Malicious (or Suspicious) boards to quantify the PDN changes due to malicious modifications. 
% For the $FD'$ histograms in this paper, without specifically pointing out, the $FD'$ is with respect to the Reference boards, where we calculate the intra-$FD'$ for all the Reference-Reference board pairs and mark them as Reference (e.g., blue bars in Fig.~\ref{fig:arduinodue_results}(b)), and we mark the inter-$FD'$ of all Reference-Malicious board pairs as Malicious (red bars in Fig.~\ref{fig:arduinodue_results}(b)).
% For PUFs, there are two important security metrics: stability (i.e., intra-HD) and uniqueness (i.e., inter-HD). 
% The former represents the ability to generate the same output for the same input under different environmental conditions, and the latter is the distinguishability of a PUF with respect to other instances. 
Here we formulate that stability (i.e., the intra-$FD'$) is the ability of two instances of the same PCB design to generate the same impedance profiles under process variation. It is calculated as $FD'(B_1^{(g)},B_2^{(g)})$, where both $B_1^{(g)}$ and $B_2^{(g)}$ are genuine boards.
The uniqueness (i.e., the inter-$FD'$) is the distinguishability of the impedance profiles of a PCB design with respect to other PCB designs. Inter-$FD'$ is calculated as $FD'(B_1^{(g)},B_2^{(a)})$, where $B_1^{(g)}$ is the genuine board, and $B_2^{(a)}$ is the board with an anomaly (i.e., malicious modifications or counterfeits).
In this paper, the $FD'$s of board pairs are plotted as histograms to show the detection performance. If the intra-$FD'$ and inter-$FD'$ are separable, then malicious boards can be identified without false positives/negatives. 
Although an ideal $FD'$ histogram should exhibit zero intra-$FD'$ and infinite inter-$FD'$, real measurements often display a more nuanced distribution.

% For the $FD'$ histograms in this paper, without specifically pointing out, the $FD'$ is with respect to the Reference boards, where we calculate the intra-$FD'$ for all Reference-Reference board pairs  and mark them as Reference (e.g., blue bars in Fig.~\ref{fig:arduinodue_results}(b)), and we mark the inter-$FD'$ of all Reference-Malicious board pairs as Malicious (red bars in Fig.~\ref{fig:arduinodue_results}(b)). An ideal $FD'$ histogram would be zero intra-$FD'$ and infinite inter-$FD'$. 

Further, we propose Frechet distance-based classification and anomaly detection methods. For anomaly detection, intra-$FD'$s are calculated for each pair of genuine boards, and the statistical boundary (e.g., $\mu$$+$$3\sigma$) of intra-$FD'$ is set as the threshold. Then the $FD'$s between the test board and the training boards are calculated and compared with the threshold to determine if an anomaly exists. 
For classification, we adopt the modified K-Nearest Neighbor (KNN) algorithm as listed in ALG.~\ref{alg:FD-KNN}.
The input of the FD-KNN algorithm includes a set of labeled training boards $\{B_i,y_i\}_{i\in[1,N]}$, a test board $B$, and the number of nearest neighbors $K$.
Instead of calculating the distance of impedance profile features, we use the $FD'$ (i.e., Eqn.~(\ref{eq:fd_multi_ports})) between the test board and the training board as the distance metric. 
The output label $y$ is decided through majority voting the labels of $K$ nearest neighbors.

\input{doc/FD_algorithm}
\subsection{Anomaly Detection} \label{sec:anmly2pdn}
%Measurement Methodology
Herein we analyze the impacts of board-level anomalies on the PDN. These are the source of \DesName{}'s high sensitivity and extensive coverage in detecting anomalies.

\smallskip
\noindent\textbf{PCB Hardware Trojan.} Board-level hardware Trojans are often implemented by adding, removing, or altering discrete components (e.g., components or programmable chips), and they may introduce large deviations in the PDN parameters. Our \DesName{} framework detects such Trojans by directly measuring changes in the PDN impedance profile. It is common for the pins of the hardware Trojan components to connect to the PDN to be powered on, which directly creates unexpected parasitic effects on the PDN. 
% Moreover, the rerouting of traces also impacts the PDN in an indirect and usually unpredictable way.
% In PCB design, the power planes fill the unused space of metal layers, so these malicious modifications may change the shape of the power planes by splitting them or changing the layers, thus indirectly influencing the board resonance. 
% In some cases, board-level hardware Trojans can maliciously snoop the debug ports to gain root privilege~\cite{exploitee2017all}.
% Even though such attacks may be elusive from the circuit schematic perspective, they affect the board resonance due to the connection/disconnection from a high-impedance pin to the PDN, and therefore can be recognized from the high-frequency region of the impedance profile. 

\smallskip
\noindent\textbf{Counterfeit or Low-Quality Electronic Components.}
% Similar to the board-level PDN, the chip-level PDN impedance profile can also be divided into two parts: the low-frequency part, caused by the chip level PDN parasitic, and the high-frequency part, caused by the package substrate resonances.
For counterfeit or low-quality chips/components, the characteristics of the PDN for both the die and package will be different from the original ones, enabling detection by \DesName{}.
% Fig.~\ref{fig:pds_methodology}(d) shows the simplified PDN structure of a chip with ball grid array (BGA) package.
Take a chip with  ball grid array (BGA) package as example. 
The package substrate has a similar structure to the PCB. The substrate consists of multiple layers connected through micro-vias to rearrange the location of pins. 
% Inside the chip package, the silicon die is connected to the package substrate through C4 bumps.
At the die level, the parasitic effects are dominated by three sources: power grid, substrate diffusion, and MOSFET gates.
Both the package and die contribute information for anomaly detection.
% There may also be on-package decoupling capacitors (OPDs) along with the die to reduce the supply noise. 
% Note that for the chips with OPDs, the different packages will contribute the most information for anamoly detection.
% At the die level, there is on-die capacitance (ODC) which is dominated by three sources: power grid metal structure, substrate diffusion, and MOSFET gates. The ODC of counterfeit chips will deviate from the genuine ones. 

\smallskip
\noindent\textbf{Board Counterfeiting/Recycling.} The impedance profile of a counterfeit/recycled PCB is partially altered by changed  specifications of the discrete components.  
In addition, due to inherent wear or aging, changes in the material characteristics of the PCB (e.g., the dielectric constant of the insulating layer) can also contribute to PDN impedance deviation and thus be detected with similar methodology.

%% file: doc/FD_algorithm.tex
\algrenewcommand\algorithmicrequire{\textbf{Input:}}
\algrenewcommand\algorithmicensure{\textbf{Output:}}

%\setlength{\textfloatsep}{0pt}

% \newlength{\textfloatsepsave} \setlength{\textfloatsepsave}{\textfloatsep} \setlength{\textfloatsep}{0pt} 
\begin{algorithm}[t]
\footnotesize
\caption{FD-based K-Nearest Neighbor Classification}
\label{alg:FD-KNN}
\begin{algorithmic}[1]
\Require $B$: test board; $K$: \# of nearest neighbor; $\{B_i,y_i\}_{i\in[1,N]}$: training boards
\Ensure $y$: class label of test board
\Function{FD-KNN} {$B,K,\{B_i,y_i\}$}
\\ \Comment{e.g., $B=\{Z_{11},Z_{12},Z_{22}\}$ when $n=2$} 
    \State Set the value of $D=[]$ 
    % \State $M=$ the number of $Z$-parameters of board $B$
    \For{$i=1$ to $N$}
    \State Calculate $d_i=FD'(B_i,B)$
     \State Append $\{y_i,d_i\}$ to $D$
    \EndFor
    \State Sort $D$ in the ascending order with respect to $d$ value
    \State Pick the first $K$ entries $\{y_j,d_j\},j=1,...,K$ from $D$
    \State $y=$ majority-voting($\{y_j,d_j\},j=1,...,K$)
\EndFunction

\end{algorithmic}
\end{algorithm}

%% file: doc/analysis.tex
In this section, we extensively analyze PDNPulse to illustrate its capability to capture board-level attacks, including PCB Trojans, chip/component counterfeits, and PCB counterfeits. We use a range of custom and COTS boards 
% (summarized in Table~\ref{tab:exp_summary}) 
to demonstrate PDNPulse's broad coverage and high sensitivity.
These PCBs cover different scales and complexities of design from 2-layer boards with tens of on-board components to 6-layer boards with hundreds of components.
% The experiments highlight one of the salient features of \DesName{}, that is it does not have hardware overheads and can be applied to legacy PCBs without any modifications.
The attacks in our evaluation are deliberately designed to model notable attacks in the real world and are also representative in terms of both stealthiness and impact.

To best illustrate the practical utility of our method, we adopt the reference designs of COTS boards. Many such designs do not provide specific PDN measuring ports. 
Thus, we employ the in-house-developed probe (see Sec.\ref{sec:experimental_setup}) to perform precise measurements. 
Without loss of generality, here we focus on one voltage domain covering the majority of the PCB, but measuring multiple domains is recommended to avoid blind points.
For the results, besides the $FD'$ histogram, we also report the  standard detection performance metrics, true positive rate (TPR), and false positive rate (FPR). An $100\%$ TPR and $0\%$ FPR would be ideal, indicating that the distributions of genuine and malicious/counterfeit boards are completely separable.

\subsection{PCB Trojan Detection}\label{sec:trojan_detection_results}
\input{doc/trojan-detection}

\subsection{Counterfeit Chip/Component Detection}\label{sec:counterfeit_chip}
\input{doc/counterfeit-chip-detection}

\subsection{Counterfeit PCB Detection}\label{sec:counterfeit_board}
\input{doc/recycling-detection}

%% file: doc/trojan-detection.tex
In this subsection, we show that %\DesName{} can effectively detect the major categories of PCB Trojans.
% Known practical PCB Trojans~\cite{zhu2021pcbench,harrison2021malicious} fall into three main categories:
% \begin{itemize}[leftmargin=*]
%     \item \textbf{Triggerable Trojans (Sec.\ref{sec:pdnpulse1v3} and \ref{sec:arduino_due_board}).} The most common Trojans are triggerable.
%     Attackers leverage small-package chips to design malicious circuits.
%     For example, chips that integrate numerous logic gates are implanted to be highly functional yet sneaky.
%     To achieve advanced attacks with complex trigger patterns or payload functions, processor chips (e.g, microncontrollers (MCU)) are commonly adopted~\cite{greenberg2019tinyspy}. 
%     \item \textbf{Always-on Trojans (Sec.\ref{sec:pdnpulse1v3}).} One notable attack is to steal sensitive information (e.g., secret keys) on the chips by inserting sampling resistors in the power rails that can perform side-channel analysis attacks~\cite{flynn2018power,wei2018know}. 
%     \item \textbf{Debug Ports Snooping (Sec.\ref{sec:pdnpulse1v3}).} Adversaries can manipulate unprotected debug ports to gain absolute privilege for performing illegal activities~\cite{exploitee2017all}.
% \end{itemize}
\DesName{} can effectively detect the three types of PCB Trojans discussed in Sec.\ref{sec:threat_model}. We fabricated genuine (i.e., Trojan-free) and malicious (i.e., Trojan-inserted) boards of two designs: a customized proof-of-concept (PoC) board and an Arduino Due board.
% A simulation model of the customized PCB is also built to show \DesName{} maintains acceptable performance facing the sneakiest modifications.
The experimental Trojans are representative of known threats and are designed to ensure the original functionality of the board. 

\begin{figure}[t]
\centering
\includegraphics[width=3.45in]{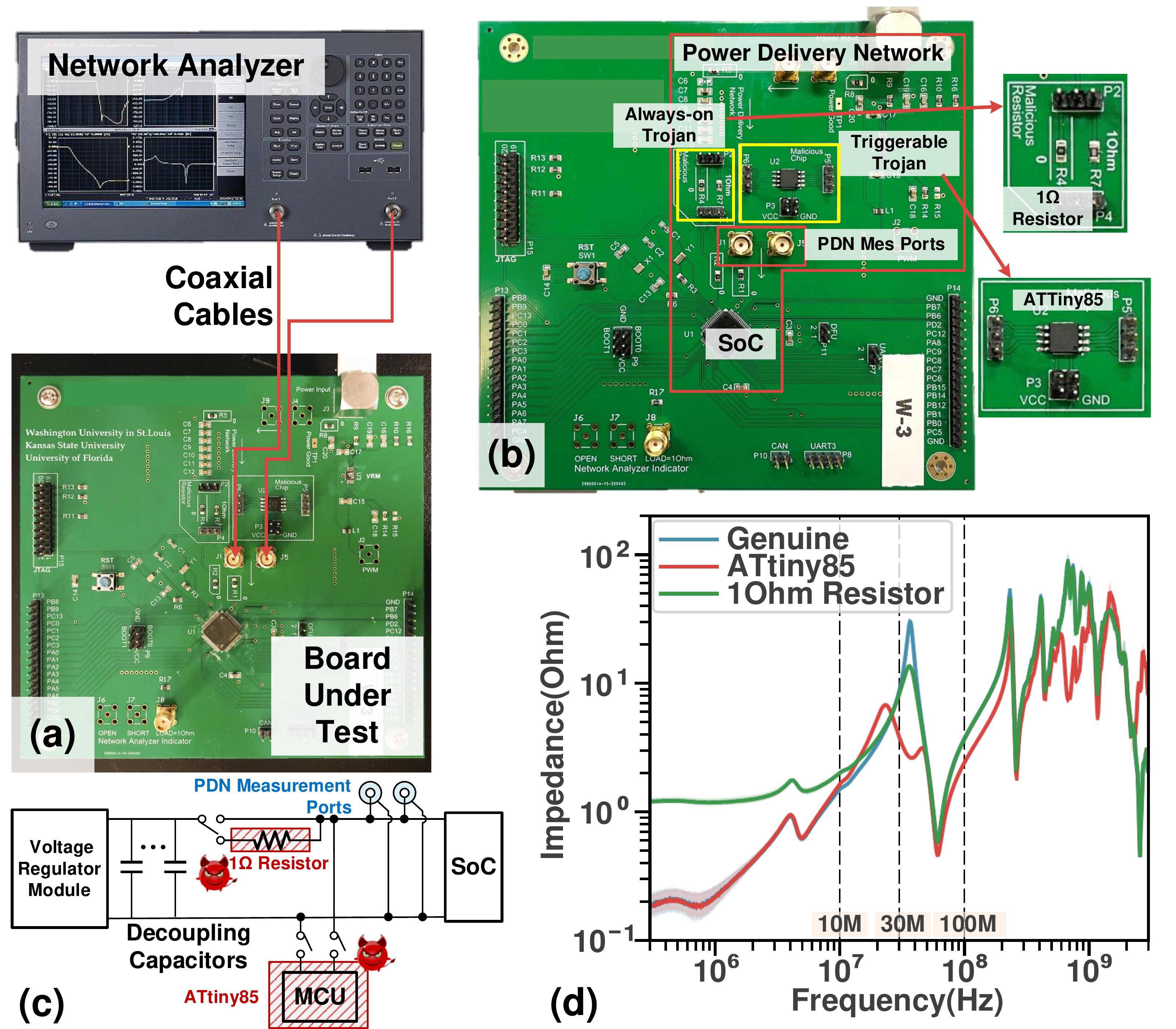}
\caption{(a) The setup for measuring the PDN impedance of customized board. (b) Layout of the customized board with Trojans inserted. (c) Simplified PDN schematic of the board. Its PDN profiles show \DesName{}'s coverage on (d) triggerable Trojan (ATtiny85) and always-on Trojan ($1\Omega$ sampling resistor).}
\label{fig:board1v3}
% \vspace{-0.1in}
\end{figure}

\subsubsection{\textbf{Coverage Evaluation with Custom PoC Boards}} \label{sec:pdnpulse1v3}

% Several respensitive hardware Trojans are inserted in the design can be connected/disconnected to the board

% 
\noindent 
We first conduct PoC experiments on a customized microprocessor development board with a relatively simple PDN to validate the coverage of \DesName{} on Trojans.

\smallskip
\noindent\textbf{Platform Description.}
Fig.~\ref{fig:board1v3} illustrates the PCB design. This is a 2-layer board containing a SoC chip and its peripheral circuit.
Its PDN follows the guideline of the SoC datasheet and has one voltage domain.
% As a proof-of-concept board, two SMA adapters connected to the PDN are added to the PCB. 
Fig.~\ref{fig:board1v3}(a) shows the experimental setup. As a custom design, we implement SMA coaxial adapters attached to the PDN, thus the VNA can directly connect to the PDN for accurate measurements. In this experiment, we fabricated and tested 4 boards. %for experimental results collections.

% \begin{figure}[t]
% \centering
% \includegraphics[width=3.0in]{figs/pdnpulse1v3_results_withROC_hardware.pdf}
% \caption{PDN profiles of customized PCB showing \DesName{}'s coverage on (a) triggerable Trojan (ATtiny85) and always-on Trojan ($1\Omega$ sampling resistor), and (b) debug ports snooping. The two booting configuration pins B0/B1 are used to mimic the snooping attacks and can be assigned to 1 by connecting to VCC.}
% \label{fig:pdnpulse1v3_results}
% \end{figure}

\smallskip
\noindent\textbf{Anomaly Description.} 
Fig.~\ref{fig:board1v3}(b) shows the three inserted Trojans. Each of them is using jumpers to turn on/off. 
% each is with jumpers to turn on/off the Trojans. 
The triggerable Trojan is an ATtiny85 MCU that performs malicious operations using a preloaded program. 
A real-world attack based on this MCU is to attack a firewall device~\cite{greenberg2019tinyspy}.
The always-on Trojan is a malicious sampling resistor enabling power side-channel analysis attacks and the value of the resistor is set to be the typical $1\Omega$~\cite{wei2018know}.
% Meanwhile, on-board jumpers for switching the boot mode of the SoC ( ``Malicious'' Booting Configuration in Fig.~\ref{fig:board1v3}(b)(c)) are utilized to mimic snooping the debug ports~\cite{exploitee2017all}, where the boot pins are similar to debug ports since both of them are composed of signal and power pins in the protocol definition.
The simplified PDN schematic with Trojans is shown in Fig.~\ref{fig:board1v3}(c).

% Compared with the probe-based measurement, the equipped SMA ports help eliminate the noise caused by the probe.
% The purpose of the above ideal experimental setup is to validate the feasibility of the proposed PDNPulse framework. 
%This ideal experimental setup is used to validate the feasibility of the proposed PDNPulse framework. 
%the test platform of the proof-of-concept results, where a customized experimental board is designed.

% Fig.~\ref{fig:board1v3}(c) shows the simplified board-level PDN schematic.
% The relatively simple development board has only one voltage domain and the PDN is composed of the VRM, decoupling capacitors, and the microprocessor.
% Note that the design of the PDN follows the normal practice of the microprocessor datasheet.
% in our experiment.

\smallskip
\noindent\textbf{Detection Results Analysis.}
% \subsubsection{Trojan Detection on Customized Boards}
We show that we can effectively detect all three Trojans,  and achieves $100\%$ TPR and $0\%$ FPR.
In this PoC experiment, the measurement is based on one port, thus we have one $Z$-parameter.
Fig.~\ref{fig:board1v3}(d) illustrate the mean $Z_{11}$ 
impedance (with $95\%$ PI shadowed\footnote{The Percentile Interval (PI) error bar is adopted here as a data spread measure, representing the spreading interval ranging from 2.5 to 97.5 percentiles~\cite{seaborn_pi}.}) when enabling different Trojans.
The blue line show the genuine impedance when none of the Trojans are connected to the circuit.

% % 
% \noindent\textbf{Triggerable Trojan Detection.}
\textit{Triggerable Trojan:}
In Fig.~\ref{fig:board1v3}(d), the red line depicts the PDN impedance profile when the ATtiny85 MCU is inserted to the circuit. This MCU's power pins are connected to the PDN and its functional pins are connected to the target signal traces.
Due to the parasitic effects of ATtiny85, an equivalent RLC network is introduced to the PDN, affecting the $R_i$, $L_j$, and $C_k$ values in Eqn.~(\ref{eq:zxy_rlc}).
Thus, a dent at $30MHz$ is observed.
Meanwhile, ATtiny85 also affects the board resonance, causing impedance profile differences above $100MHz$.

% % 
% \noindent\textbf{Always-on Trojan Detection.}
\textit{Always-on Trojan:}
The green line in Fig.~\ref{fig:board1v3}(d) depicts the PDN impedance profile when the $1\Omega$ sampling resistor is connected in series to the PDN. 
Due to the insertion of the resistor, the impedance below $10MHz$ raises from $0.2 \Omega$ to $1.2 \Omega$, and the impedance around $30MHz$ also deviates from the genuine impedance. 
Since the genuine impedance at low frequency is $0.2\Omega$, with $\pm10\%$ PCB fabrication tolerance, we can detect malicious resistors larger than $0.02\Omega$. 

% % 
% \textit{Debug Ports Snooping:}
% Fig.~\ref{fig:board1v3}(e) shows that the PDN is sensitive to snooping on the debug ports due to the board resonance, even though this attack negligibly impacts the PCB or on-board components.
% Impedance profiles of three different booting configuration are compared, since there are two booting configuration pins, namely: B0, B1, and B0+B1.
% Under each configuration, the corresponding pins (B0 and/or B1) are connected to the VCC to assign the pins logic 1.
% Since the jumper connects high-impedance pins to the much low-impedance PDN, their impact on PDN cannot be detected from the circuit perspective.
% Thus, below $100MHz$, all four profiles are overlapped.
% However, these three booting configurations greatly affect the board resonance and can be clearly recognized from the impedance changes in the high frequency range. 
% The detection sensitivity is also demonstrated since connecting one or two pins can be easily distinguished.

\begin{figure}[t]
\centering
\includegraphics[width=3.45in]{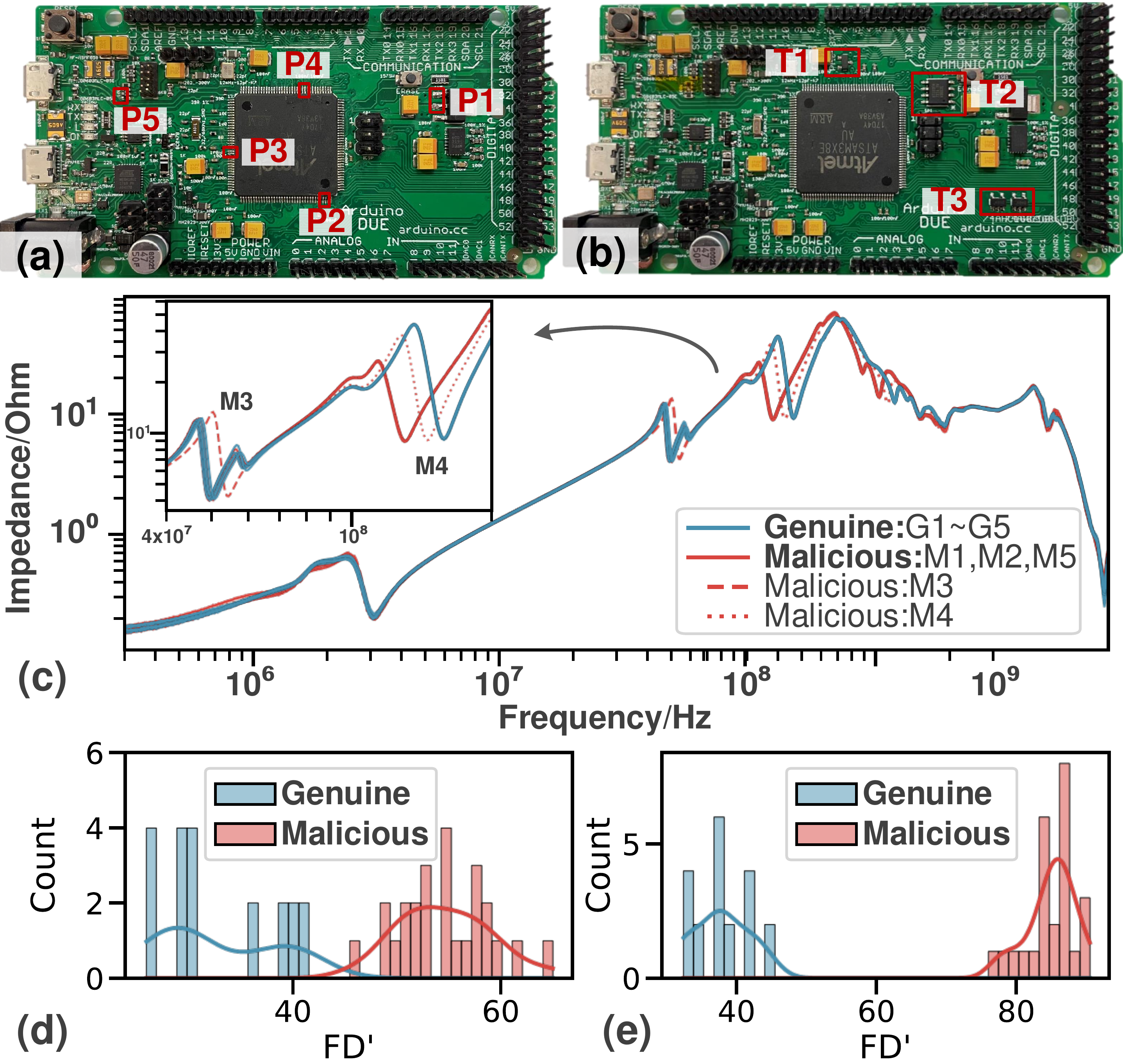}
\caption{Layouts of the (a) genuine Arduino Due board with measurement ports highlighted, and (b) the malicious board with three triggerable Trojans highlighted. (c) Impedance profiles (i.e., $Z_{55}$) of the 5 genuine boards (G1$\sim$G5) and 5 malicious boards (M1$\sim$M5), where each of the 10 curves represents one board. (d)(e) Histogram of $FD'$ of all board pairs calculated using the $Z$-parameters of (d) 3 ports (P1-P3) and (e) all 5 ports. Malicious boards can be clearly distinguished from genuine boards.}
\label{fig:arduino_due}
% \vspace{-0.1in}
\end{figure}

% \begin{figure}[t]
% \centering
% \includegraphics[width=3.1in]{figs/arduinodue_results_noZ15.pdf}
% \caption{(a) Impedance profiles (i.e., $Z_{55}$) of the 5 genuine Arduino Due boards (G1$\sim$G5) and 5 malicious boards (M1$\sim$M5), where each of the 10 curves represents the profile of one board. And the histogram of $FD'$ of all board pairs calculated using the $Z$-parameters of (b) 3 ports (P1-P3) and (c) all 5 ports. From the results, malicious boards can be easily distinguished from genuine boards.}
% \label{fig:arduinodue_results}
% \end{figure}

\subsubsection{\textbf{Triggerable Trojan on Arduino Due Boards}} \label{sec:arduino_due_board}

\noindent Herein, we validate \DesName{} can detect the foremost type of Trojans, triggerable Trojans, on complex COTS Arduino Due boards~\cite{arduino_due} using designed probes. 
% which is a commercial-level design following the industrial PCB design standards.
Three different triggerable Trojans are inserted to the PCBs during the design stage.

\smallskip
\noindent\textbf{Platform Description.}
Fig.~\ref{fig:arduino_due}(a) shows the genuine Arduino Due board.
It is a 2-layer MCU development board based on ATsam3x8e.
There are four voltage domains (5V, 3.3V, USBVCC, and XVCC), supporting more than 50 on-board components.
In this experiment, we focus on the 3.3V voltage domain since it supplies power to most on-board components.
% We abstract the PDN 
The PDN is abstracted into a 5-port network and the locations of PDN measurement ports are highlighted in Fig.~\ref{fig:arduino_due}(a).
The ports are ensured to diversely distribute on the PCB and not on or close to the capacitors.
% For instance, since there are no capacitors near port P5, thus the self-impedance $Z_{55}$ is expected to reveal the changes across the board.
We fabricate 5 genuine Arduino Due boards and 5 malicious ones for testing.

\smallskip
\noindent\textbf{Anomaly Description.}
Fig.~\ref{fig:arduino_due}(b) highlights the three Trojans, which are carefully designed to be impactful while sneaky. 
T1 is a small-package MOSFET chip for leaking information through a LED. The LED was used for indicating the output pulse-width modulation signal.
At T2, a maliciously programmed MCU ATtiny102F is implemented. Attackers can send messages through UART to the board and trigger ATtiny102F to erase the on-board Flash memory.
T3 is based on two 74xx XOR logic chips with small packages, which can be triggered by the software and then crash the system.
The readers are referred to~\cite{zhu2021pcbench} for more details of the Trojans.

\smallskip
\noindent\textbf{Detection Results Analysis.}
% \subsection{Trojan Detection on Arduino Due Boards}
Fig.~\ref{fig:arduino_due}(c)-(e) show the robustness and effectiveness of our framework. We also find that \DesName{} is sensitive to changes in PCB design while resistant to process variation.
We obtain 15 $Z$-parameters for each of the 10 boards, then calculate $FD'$ for $5\times4$$=$$20$ genuine-genuine pairs and $5$$\times$$5$$=$$25$ genuine-malicious pairs.

Fig.~\ref{fig:arduino_due}(c) shows that due to the Trojans, the malicious boards can be distinguished from genuine boards with process variation, where we exhibit one of the measured impedance profiles, $Z_{55}$, and plot the profiles of all 10 boards.

In Fig.~\ref{fig:arduino_due}(d)(e), histograms of $FD'$ demonstrate the effectiveness of FD-based detection algorithm and the benefits of multi-port detection.
For the $FD'$ histograms in this paper, the $FD'$ is with respect to the genuine boards, where we calculate the intra-$FD'$ for all genuine-genuine board pairs and mark them as Genuine (blue bars in Fig.~\ref{fig:arduino_due}(d)), and we mark the inter-$FD'$ of all genuine-malicious board pairs as Malicious (red bars in Fig.~\ref{fig:arduino_due}(d)). 
% Here, the L1 norm is used in $FD'$ calculation.
Fig.~\ref{fig:arduino_due}(d) is based on 3-port detection, where only ports P1-P3 (with 6 $Z$-parameters) are utilized.
The gap between the intra-$FD'$ and inter-$FD'$ implies that a threshold can be set to identify potential boards. 
Then Fig.~\ref{fig:arduino_due}(e) shows the results leveraging all 5 ports (with 15 $Z$-parameters).
The gap increases, making the detection more resistant to PCB tolerance and sensitive to Trojans.
In both 3-port and 5-port detection, we achieve $100\%$ TPR and $0\%$ FPR.

We further show that \DesName{} maintains its sensitivity on complex COTS boards.
We remove one chip at T3 on the M4 board (the Malicious board with index 4) during the testing for ease of debugging. 
The profile of M4 board is different from other Malicious boards, as illustrated in the inset of Fig.~\ref{fig:arduino_due}(c).
The package of the removed chip is smaller than any other digital chip on this board. 
Since the parasitic effects are correlated with the chip's physical size~\cite{kim2010chip}, this result indicates the sensitivity of \DesName{}.

% It is also worth noting that, besides M4, the impedance profiles of M3 board are surprisingly also different from the others. 
% Although we do not find any difference between M3 and other Malicious boards in appearance, we believe this is not due to process variation since the difference is in a specific frequency band and is consistent among multiple $Z$-parameters.
% It may be due to the flaws of the board. 
% }

% High low-frequency impedance in M3-P11,P12,P13(compared with malicious boards), P14(compared with malicious boards), P15 (most obvious)
% P22 low impedance
% P24 low frequency, low impedance
% P25 shift
% P33 high impedance, low frequency
% P34 mid frequency, low impedance
% P35 shift
% P45 shift
% P55 shift
% not able to recognize the modifications

%% file: doc/counterfeit-chip-detection.tex
In this subsection, we demonstrate the desirable performance and sensitivity of \DesName{} in detecting both counterfeit chips and components.
% while existing works (e.g., JTAG-based method~\cite{paul2021silverin}) cannot easily cover them simultaneously.
While, existing works may fail to detect both types since they typically utilize specific design features that are not compatible with both chips and components (e.g., JTAG, which is only available in chips~\cite{paul2021silverin}).
Besides, the labels (i.e, markings) of chips/components can be removed or occluded (e.g., by EM shields), invalidating imaging/visual inspection. We overcome these limitations by utilizing unified PDN electrical properties.
Here we perform detection on three platforms: PYNQ-Z1 and PYNQ-Z2 FGPA development board, and MSI H310M computer motherboard.
The target chips/components are in various packages to show \DesName{}'s coverage.

\begin{figure}[t]
\centering
\includegraphics[width=3.45in]{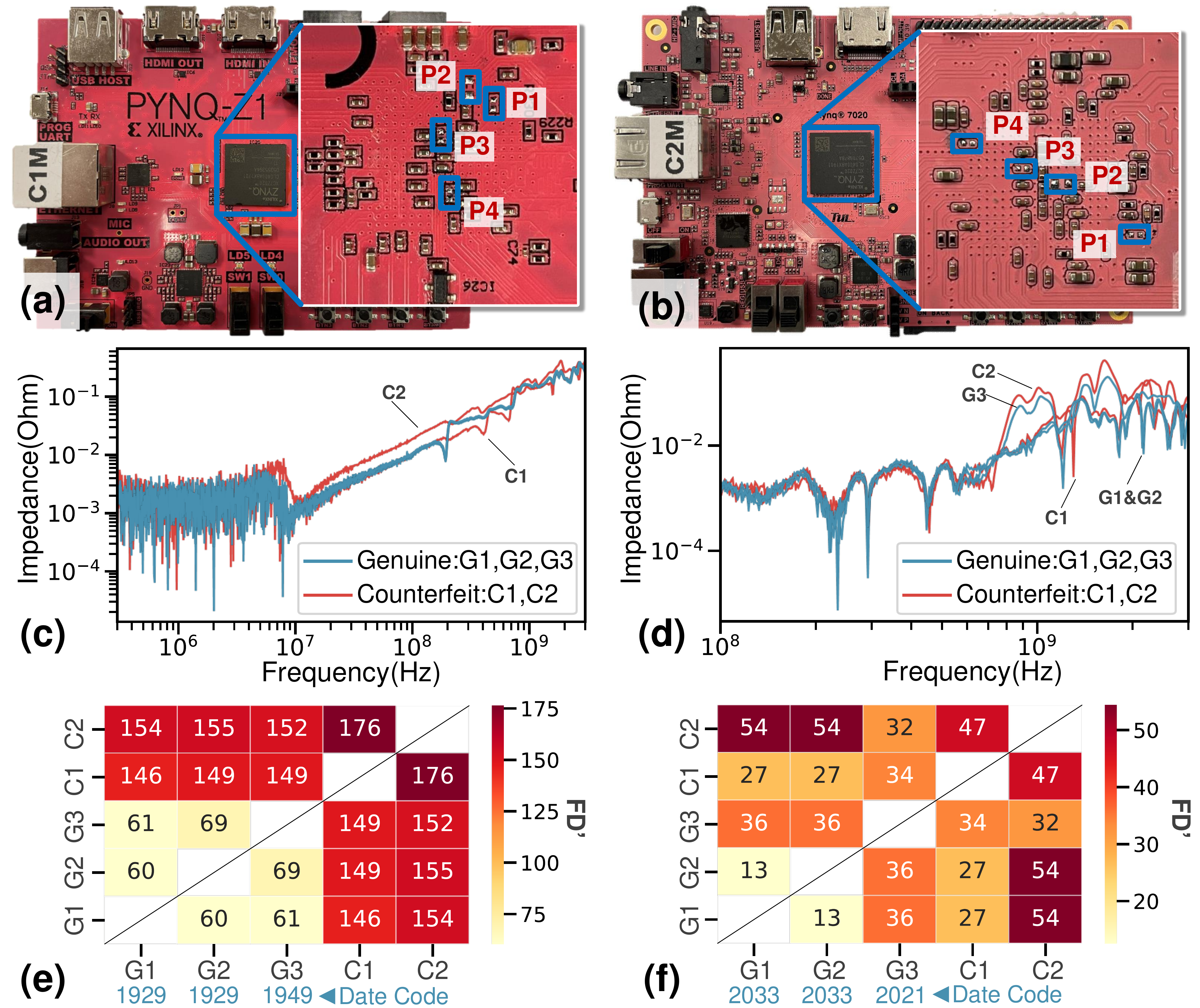}
\caption{The layouts (a)(c)(e) PYNQ-Z1 and (b)(d)(f) PYNQ-Z2 boards, with results showing \DesName{}'s effectiveness in detecting counterfeit chips. (c)(d) The $Z_{12}$ profiles of 5 boards, with each curve representing one board. (e)(f) Heat maps show $FD'$ based on all $Z$-parameters, with FPGA chip date codes highlighted. }
\label{fig:pynq_results}
% \vspace{-0.1in}
\end{figure}

\subsubsection{\textbf{Counterfeit Chip on PYNQ Boards}} \label{sec:pynq}
\noindent 
% \DesName{} can detect counterfeit chips in various packages including BGA package and quad flat package (QFP), which are intentionally chosen to reflect the variety of package forms used in the chips.
% We detect counterfeit chips in BGA package on two platforms, PYNQ-Z1 and PYNQ-Z2 boards, then conduct experiments on Trezor One board that is based on a MCU chip in a QFP. Here we mainly introduce details for detecting PYNQ boards, and evaluate the Trezor One board in Appx.~\ref{apx:trezor_one}, where similar experiments are conducted.
% The results in Appx.~\ref{apx:trezor_one} show that we can effectively ($100\%$ TPR, $0\%$ NPR) identify the counterfeit chip in a QFP.
% \DesName{} can detect counterfeit chips in various packages including BGA package and quad flat package (QFP), which are intentionally chosen to reflect the variety of package forms used in the chips.
We detect counterfeit chips on two platforms, PYNQ-Z1 and PYNQ-Z2 boards. 

\smallskip
\noindent\textbf{Platform Description.}
Fig.~\ref{fig:pynq_results}(a) and (b) highlight the measured ports on both PYQN boards.
These two boards are FPGA development boards based on the Xilinx XC7Z020 FPGA.
The boards have 6 layers and more than 100 on-board components. 
In this experiment, we purchase 5 PYNQ-Z1 boards and 5 PYNQ-Z2 boards.
The PDNs are elaborately designed, including more than 15 voltage domains. The FPGA chip is supplied by 4 voltage domains (3.3V, 1.0V, 1.8V, and 1.5V). 
% The FPGA chips are surrounded by sufficient capacitors to reduce the supply noise. One challenge here is that since the chip-level PDN is in parallel with the board-level PDN, the low-impedance board-level PDN may dominate the measurement and make any changes of chip-level PDN negligible. 
% To solve this issue, we propose to perform inter-domain detection and specifically focus on the coupling effects among different voltage domains.
% These coupling effects at the chip level are much higher than the PCB level and thus reduce the interference from the board-level PDN.
% Since the different domains do not have direct electrical connections, the transfer impedance across the two voltage domains is quite small at low frequencies.
% However, the transfer impedance at high frequencies can deviate from each other, which mainly result from differences in the package of chips, facilitating counterfeit chip detection.
We select the ports from these 4 voltage domains (at the bottom side of the FPGA), and abstract the PDN into a 4-port network with each port for one domain.
% A moving average filter with 24 points window width is applied correspondingly when calculating $FD'$ to filter out the interference of low-frequency parts.

% from different distributors such as Digi-Key, Amazon, and Mouser Electronics.

\smallskip
\noindent\textbf{Anomaly Description.}
To mimic chip counterfeiting, we replace the original FPGA chips of 2 PYNQ-Z1 boards and 2 PYNQ-Z2 boards with recycled ones considering that chip recycling is the main source for chip counterfeiting. Note that these recycled chips may not be authentic before recycling.
% }

\smallskip
\noindent\textbf{Detection Results Analysis.}
Fig.~\ref{fig:pynq_results}(c)-(f) illustrate that the counterfeit FGPA chips are distinguishable from genuine ones. 
% The latter indicates that the process variation of the chip itself should be considered when establishing genuine boards. 
% We present the results of PYNQ-Z1 in  Fig.~\ref{fig:pynq_results}(c)(e) and the results of PYNQ-Z2 in Fig.~\ref{fig:pynq_results}(d)(f).
Since there are 4 ports in 4 voltage domains, 10 $Z$-parameters are available. 
We collect 6 non-diagonal $Z$-parameters (e.g., $Z_{12}$, $Z_{24}$) for each board to emphasize the coupling effects between voltage domains. 
% We obtain $FD'$ for each board pair.
In Fig.~\ref{fig:pynq_results}(c)(d), one of the $Z$-parameters, $Z_{12}$ of all boards are plotted to show the differences between genuine boards and counterfeit ones.
% We also highlight the difference due to different board distributors (i.e., Amazon and Mouser) in Fig.~\ref{fig:pynq_results}(d), where the chip batch number is related to the board distributor. 
% In both figures, since the different domains do not have direct electrical connections, the inter voltage domain transfer impedance is quite small at low frequencies.
% However, due to the crosstalk, the transfer impedances of the boards at high frequencies ($\sim 100MHz$) can deviate from each other, which mainly result from differences in the package of chips, facilitating counterfeit chip detection.

Then, in Fig.~\ref{fig:pynq_results}(e)(f), we use $FD'$ matrix to quantitatively validate the results in Fig.~\ref{fig:pynq_results}(c)(d) by considering all measured $Z$-parameters and analyzing the relationship between FPGA chip date codes~\cite{xilinxdeviceguide}.
% In the figures, we follow such a naming scheme for each board: [board type (C/G) + index number + distributor(A/M/D)], where C means counterfeit, G means genuine, and A, M, and D represent Amazon, Mouser, and Digi-Key, respectively.
In Fig.~\ref{fig:pynq_results}(e), the genuine boards have low $FD'$ with each other while having high $FD'$ with the counterfeit boards. Thus, a threshold can be set to identify counterfeit boards with $100\%$ TPR and $0\%$ FPR.
% Later, we confirm that the C2D board contains a short in its PDN, probably due to the counterfeit chip. For the same reason, the C2D board has the highest FD' among all boards. 
Compared with PYNQ-Z1, the differences between PYNQ-Z2 boards with different FPGA date codes are more obvious.
Interestingly, as shown in Fig.~\ref{fig:pynq_results}(f), the $FD'$ between G1 and G3 or between G2 and G3 are even higher than the $FD'$ between G1 and C1.
We find that both G1 and G2 are manufactured at 2033 (the 33th week of 2020), while G3 is manufactured at 2021 (the 21st week of 2020).
We also see this relationship in Fig.~\ref{fig:pynq_results}(e), where G1 (date code: 1929) is closer to G2 (1929) compared to G3 (1949).
For PYNQ-Z2, single threshold-based detection yields false positives/negatives. As will be discussed in Sec.\ref{sec:discussion}, these faults can be avoided by modeling multiple batches and applying the FD-KNN algorithm.
% Therefore, the FPGA chips with the same date code have similar impedance profiles while different date codes have dissimilar impedance profiles. 
% This indicates the process variation between chip batches should be considered building the golden model to avoid false positives.

To further analyze the impact of de-/re-soldering the chips on anomaly detection, we de-solder the FPGAs chips on all PYNQ-Z1 boards (i.e., including the genuine ones and counterfeit ones) and then re-solder the chips. 
For each PYNQ-Z1 board, the impedance profiles are measured again, and the $FD'$ between the newly and previously measured impedance profiles are calculated.
The $FD'$ ranges from 2.7 to 11.4, indicating that if we set the anomaly detection threshold at 100 (see Fig.~\ref{fig:pynq_results}(e)), no false positive nor false negative will be induced due to the de-/re-soldering operations.

\subsubsection{\textbf{Counterfeit Component on Motherboards}}\label{sec:motherboard}
\noindent We further show both \DesName{}'s effectiveness in detecting counterfeit components and its scalability for assuring the security of large-scale PCB designs.
The high sensitivity of \DesName{} for detecting counterfeit chips/components is also validated. 

\begin{figure}[t]
\centering
\includegraphics[width=3.45in]{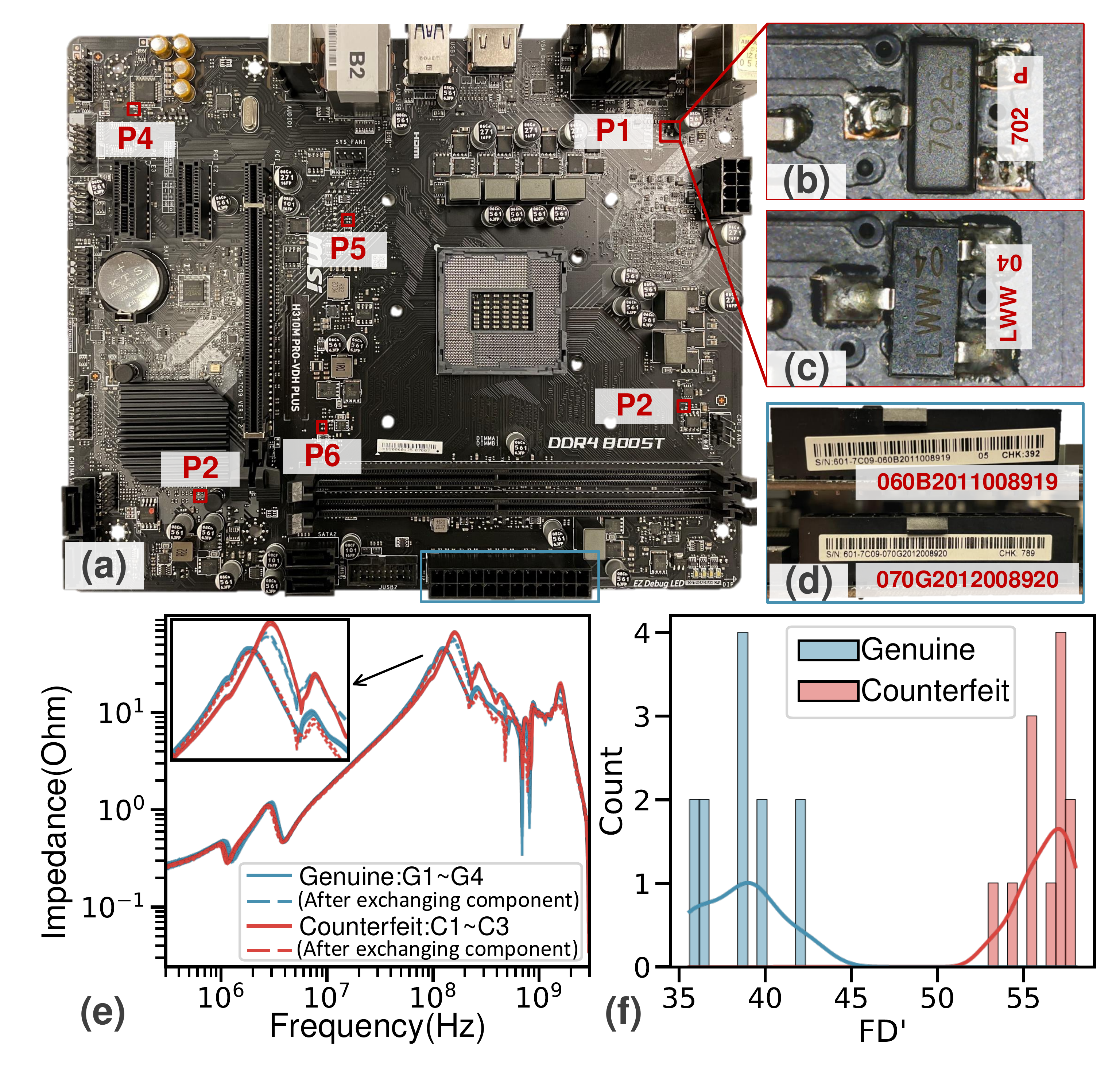}
\caption{(a) The layout of MSI H310M computer motherboard with measurement ports highlighted. (b)(c) The labels of the transistor at P1 on boards fabricated by two production lines are different. (d) The two production lines have different S/N codes. (e) Seven $Z_{11}$ profiles including 4 genuine and 3 counterfeit boards, where the boards from one production line are deliberately regarded as counterfeits. The 4 dashed lines 
(2 genuine boards with counterfeit transistor, and 2 counterfeit boards with genuine transistor)
confirm the difference between the two types of boards. (f) The $FD'$ histogram based on all $Z$-parameters shows detection effectiveness.}
\label{fig:motherboard_layout}
% \vspace{-0.1in}
\end{figure}

\smallskip
\noindent\textbf{Platform Description.}
Fig.~\ref{fig:motherboard_layout}(a) shows the selected large-scale design, MSI H310M computer motherboard, and PDN measurement ports.
The board contains 6 layers and hundreds of components.
On the motherboard, there exist more than 30 voltage domains and the supply voltage ranges from 1V to 12V with different specifications.
For most voltage domains, the supplied components  usually concentrate in a small area of the board.
In this experiment, we focus on the 3.3V voltage domain and abstract the PDN to a 6-port network.
The components supplied by 3.3V voltage domain spread across the board, thus we can evaluate the board with one domain. 
We purchase 4 and 3 brand-new motherboards fabricated by two production lines (confirmed by checking the S/N code shown in Fig.~\ref{fig:motherboard_layout}(d)) for PDN measurements.
% 
% \begin{figure}[t]
% \centering
% \includegraphics[width=3.3in]{figs/motherboard_results_chipswitched.pdf}
% \caption{(a) Seven impedance profiles of $Z_{11}$ including 4 genuine and 3 counterfeit motherboards, where the boards from one production line are deliberately regarded as counterfeit to mimic counterfeiting attacks. The 4 dashed lines 
% (2 genuine boards with counterfeit transistor, and 2 counterfeit boards with genuine transistor)
% validate the difference between the two types of boards. (b) The histogram of $FD'$ of all board pairs calculated using $Z$-parameters of 6 ports to quantitatively show detection effectiveness.}
% \label{fig:motherboard_results}
% \end{figure}
% 
% 
We further show both \DesName{}'s effectiveness in detecting counterfeit components and its scalability for assuring the security of large-scale PCB designs.
% Its high sensitivity for detecting counterfeit chips/components is also illustrated. 

% 

\smallskip
\noindent\textbf{Anomaly Description.}
While both boards are legitimate versions, the BOMs of the two production lines are different. Thus, to mimic component counterfeit, the boards from one production line are regarded as genuine and the other production line is treated as counterfeit. 
% We apply \DesName{} to compare the quality of on-board components from two production lines. 
Based on the PDN measurements, after carefully examining the two motherboards, we notice that only the transistor at P1 (see Fig.~\ref{fig:motherboard_layout}(a)) for power management is different in both boards (see Fig.~\ref{fig:motherboard_layout}(b) and (c)).
This transistor is thus utilized as an instance of mimicking component counterfeiting attacks.

\smallskip
\noindent\textbf{Detection Results Analysis.}
% \subsection{Computer Motherboard from Distributors}
Fig.~\ref{fig:motherboard_layout}(e) illustrates the impacts on PDN impedance profile due to the transistor at P1. With 6 ports, we measure 21 $Z$-parameters for each board.
Here, the $Z_{11}$ of all 7 motherboards are plotted, where the differences between the 4 genuine and 3 counterfeit boards are distinguishable.
By examining other $Z$-parameters, the differences are observable mainly in the $Z$-parameters related to P1.
Since the two boards have different transistor models at P1, even though the two transistors have the same package, the impedance profiles are different from each other.
To confirm this finding, we exchange this transistor among four boards (2 from genuine boards, 2 from counterfeit boards). As shown in Fig.~\ref{fig:motherboard_layout}(e), the 2 blue dashed lines are the genuine boards with transistors from counterfeit boards, and the 2 red dashed lines are counterfeit boards with genuine transistors. We observe that the impedance profiles of the modified boards match with the other type of boards, meaning that the transistor at P1 causes the differences between profiles.

We further use the $FD'$ based on all 21 $Z$-parameters to quantitatively show \DesName{} effectively detects the counterfeit component. There are total 12 genuine-genuine pairs and 12 genuine-counterfeit pairs.
Fig.~\ref{fig:motherboard_layout}(f) shows the histogram of board pairs. The TPR and FPR are $100\%$ and $0\%$, respectively.
% Our experimental results validate that components of different qualities will impact the overall PDN performance (and the overall board performance). 
% Therefor, \DesName{} has a good scalability for complex designs such as motherboards.
Since we detect a counterfeit component in a small package on a relatively large-scale PCB, the results not only show \DesName{}'s scalability for complex designs such as motherboards, but also indicate its acceptable sensitivity in counterfeit chip/component detection.
%and has the ability of detecting both counterfeit chips and counterfeit components.

%% file: doc/recycling-detection.tex
\begin{figure}[t]
\centering
\includegraphics[width=3.45in]{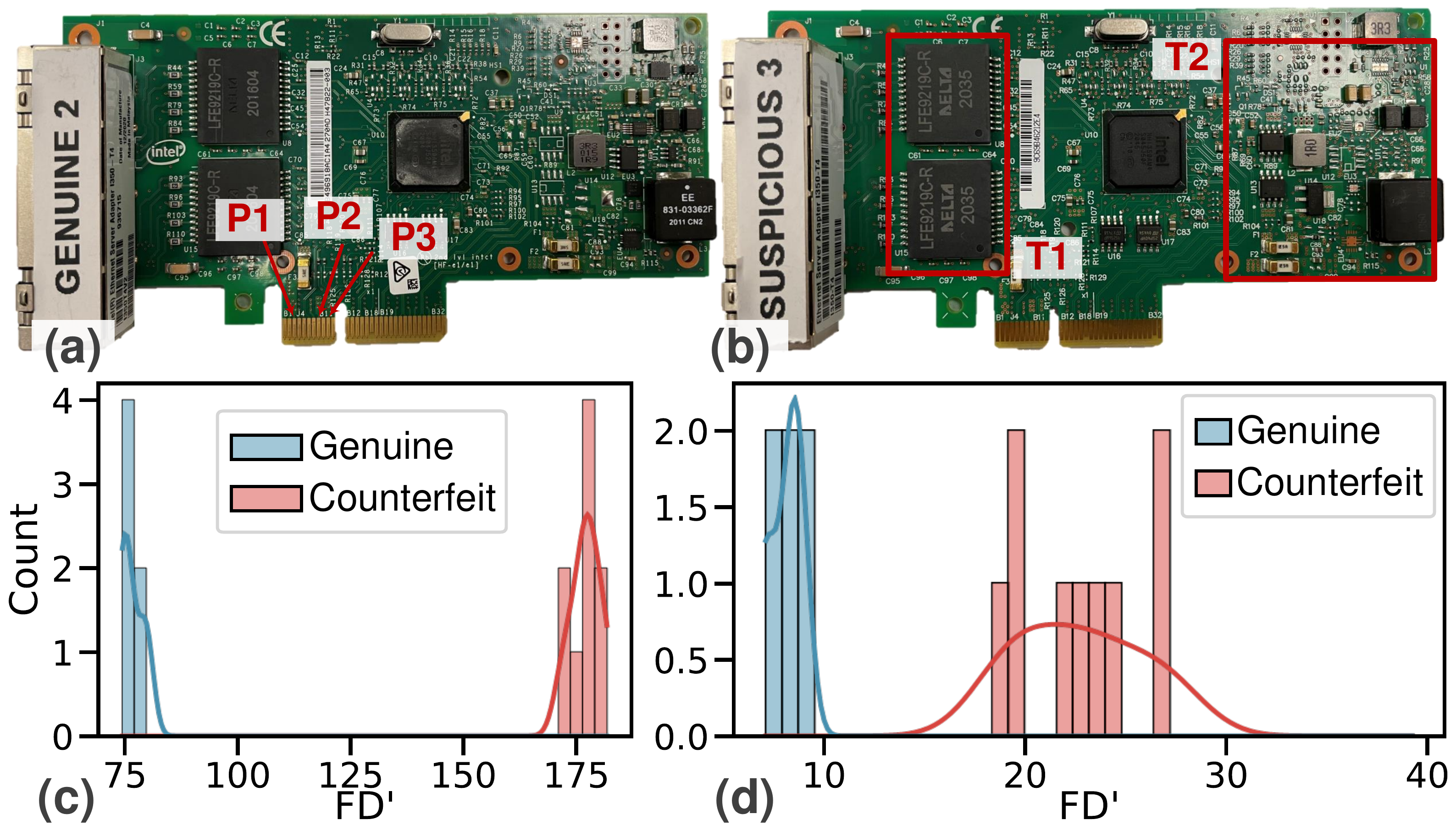}
\caption{(a) Genuine Intel I350-T4 board with measurement ports highlighted. (b) Counterfeit board with suspicious areas marked. The $FD'$ histograms of (c) direct detection and (d) cross-board detection, showing \DesName{} can detect board imitations.}
% \vspace{-0.1in}
\label{fig:network_adapter_layout}
\end{figure}

In this subsection, we show that \DesName{} can also detect PCB counterfeit with varying degrees of sneakiness.
% Real-world counterfeit PCBs typically have three types: 
% %according to the sneakiness:
% \begin{itemize}[leftmargin=*]
%     \item \textbf{Imitating  (Sec~\ref{sec:i350t4}).} 
%     Attackers have complete access to the PCB design resources. However, for higher profits, they typically replace parts of the original circuit with a low-standard design.
%     % Attackers have access to only part of the PCB design resources. 
%     The fabricated counterfeit boards are thus different from the original PCBs, which is sometimes directly observable from the board layout.
%     Still, they can remain undetected since the boards are usually inside the products (e.g., servers), preventing imaging inspection.
%     \item \textbf{Cloning (Sec~\ref{sec:arduino_uno}).} Adversaries also have all PCB design resources. They can fabricate the board with the same layout while embedding counterfeit or low-quality components. This type of counterfeit can be quite difficult to visually distinguish from a genuine board.
%     \item \textbf{Recycling (Sec~\ref{sec:arduino_uno}).} Genuine PCBs are purchased, depreciated, recycled, and sold as brand-new products. This is the sneakiest type of counterfeit since recycled boards can have the same appearance as genuine PCBs.
% \end{itemize}
% We show that \DesName{} can detect all three types of counterfeits.
% \DesName{} captures changes in specifications caused by using counterfeit and/or low-quality discrete components and materials on the PCB.
% Thus, we can use a similar methodology applied for detecting Trojans and counterfeit chips/components. 
Two PCB designs are selected, an Intel I350-T4 Ethernet adapter and an Arduino Uno board, covering the two types of PCB counterfeit.

\subsubsection{\textbf{Imitation Network Adapter Boards}}\label{sec:i350t4}
% % 
We demonstrate the capability of \DesName{} in detecting an imitation Intel I350-T4 Ethernet adapter.
% \dean{It seems like you're trying to make a distinction between two imitated boards then. Shouldn't it be between a genuine board and its imitation? So, potential edit: "We demonstrate the capability of \DesName{} in detecting an imitation Intel I350-T4 Ethernet adapter." Unless you're saying it's an imitation of an imitation?} \huifeng{They are the same.}
This imitation is a real-world attack and is well documented~\cite{comparison_intel_i350}.
We also illustrate \DesName{} can perform cross-board detection, where imaging inspection may not be available.
% We then apply PDNPulse to a real-world counterfeit board, the Intel I350-T4 Ethernet server adapter, and its counterfeit board.

\smallskip
\noindent\textbf{Platform Description.} Intel I350-T4 is a network adapter board based on the Intel I350 processor.
% In this experiment, we first rely on the power pins on the PCI-E connector and perform the cross-domain detection.
There are three voltage domains on PCI-E connector as well as the adapter board, 12V, 3.3V, and 3.3Vaux.
Correspondingly, we abstract the PDN to a 3-port network for direct detection and the locations of these ports are shown in Fig.~\ref{fig:network_adapter_layout}(a).
For cross-board detection, we plug the network adapter into the MSI H310M PRO-VDH PLUS computer motherboard. The probe is attached to the PCI-E 12V power pin on the motherboard.
Note that here the computer motherboard is not powered.

\smallskip
\noindent\textbf{Anomaly Description.}
We purchase 3 genuine boards and 3 counterfeit ones, as shown in Fig.~\ref{fig:network_adapter_layout}(a) and (b).
We confirm that the counterfeit boards are imitated boards for two reasons. 
First, the word ``Delta'' on the Delta Ethernet transformers (T1 in Fig.~\ref{fig:network_adapter_layout}(b)) should be embossed on the chip.
Second, the peripheral circuit (T2 in Fig.~\ref{fig:network_adapter_layout}(b)) is replaced with a low-standard design.
As reported in~\cite{comparison_intel_i350}, the counterfeit boards are equipped with low-quality chips and components such that these boards will probably fail within one year.

% \begin{figure}[t]
% \centering
% \includegraphics[width=3in]{figs/network_adapter_results_fd.pdf}
% \caption{The $FD'$ between genuine and counterfeit adapter boards when performing (a) direct detection and (b) cross-board detection using the computer motherboard, showing \DesName{}'s capability on detecting board imitation. Open is measured when no adapter board is connected to the motherboard.}
% \label{fig:network_adapter_results}
% \end{figure}

\smallskip
\noindent\textbf{Detection Results Analysis.}
% \subsection{Counterfeit Network Adapter Boards}
Fig.~\ref{fig:network_adapter_layout}(c)(d) show \DesName{}'s effectiveness of both direct and cross-board detection. We obtain $FD'$ for 6 genuine-genuine pairs, and 9 genuine-counterfeit pairs.
In Fig.~\ref{fig:network_adapter_layout}(c), the $FD'$ are calculated based on all 6 $Z$-parameters (3 ports).
Even though the two boards have the same PCB layout, due to the usage of different on-board components, the intra- and inter-$FD'$ are significantly different. 
Fig.~\ref{fig:network_adapter_layout}(d) shows the results of cross-board detection, where $FD'$ is based on one $Z$-parameter.
% Besides genuine and counterfeit, the green bars (i.e., Open) are configured such that no board is connected to the motherboard's PCI-E.
%no board is connected to the PCI-E. 
In cross-board detection, the impedance profiles of network adapters are distorted by the PDN of the motherboard and the parasitics of PCI-E connector, which reduces the inter-$FD'$. 
However, we still can identify the counterfeit boards. We have $100\%$ TPR and $0\%$ FPR in both types of detection.

\subsubsection{\textbf{Cloned Arduino Uno Boards}}\label{sec:arduino_uno}
% \huifeng{Threat Model}
% 
We then demonstrate \DesName{} on cloning detection by experimenting with Arduino Uno boards~\cite{ArduinoUno}, which are popular in the market for their low cost.

\begin{figure}[t]
\centering
\includegraphics[width=3.45in]{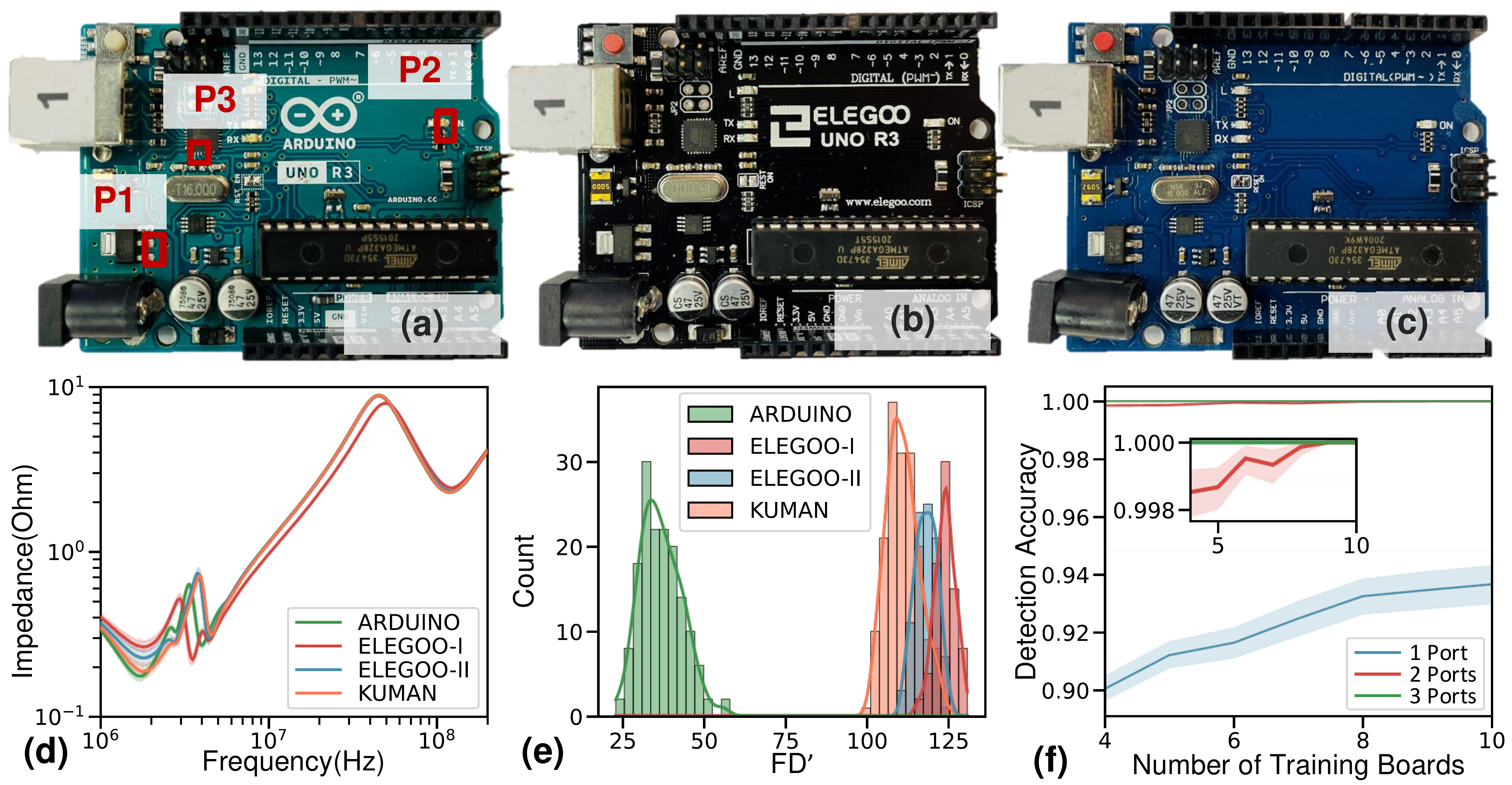}
\caption{Arduino Uno boards from three vendors: (a) official Arduino, (b) Elegoo, and (c) Kuman. We treat the official Arduino as genuine, and the other two as counterfeit to mimic cloned PCBs. (d) The mean $Z_{33}$ profiles  (with $95\%$ PI) of Arduino boards. (e) Histogram of $FD'$ based on all $Z$-parameters, where the boards can be clearly identified as genuine or counterfeit. (f) Results of FD-3NN classification for different numbers of measurement ports.}
\label{fig:arduino_uno_layout}
% \vspace{-0.1in}
\end{figure}

\smallskip
\noindent\textbf{Platform Description.}
Arduino Uno is an open-source MCU development board based on the ATmega328P. It has been fabricated and sold by many manufacturers, which serves as an excellent example of cloning attacks.
Fig.~\ref{fig:arduino_uno_layout}(a) shows the locations of the three measurement ports. We measure the 5V voltage domain which is the main supply voltage of the PCB and abstract the PDN to a 3-port network. 

% \begin{figure}[t]
% \centering
% \includegraphics[width=3in]{figs/arduino_uno_results_new_fd.pdf}
% \caption{(a) Comparison of mean $Z_{33}$ profiles  (with $95\%$CI) of Arduino Uno boards from different vendors. (b) The histogram of $FD'$ based on all $Z$-parameters, where the boards can be successfully identified as genuine or counterfeit boards. (c) Results of FD-3NN classification under different numbers of measurement ports. (d) The mean $Z_{22}$ profiles (with $95\%$CI) of the genuine and recycled boards showing the effectiveness of recycling detection.}
% \label{fig:arduino_uno_results}
% \end{figure}

\smallskip
\noindent\textbf{Anomaly Description.}
% It is popular for its low cost, so many manufacturers have fabricated the board.
% Fig.~\ref{fig:arduino_uno_layout} shows the Arduino Uno boards manufactured by different vendors for cloning detection. 
%The Arduino Uno is a microcontroller development board based on the ATmega328P and developed by Arduino.cc.
%It is an open-source hardware design where the schematic, bill-of-material (BOM), and layout are public.
%Therefore many different vendors are able to build their Arduino Uno boards.
% In our experiment, we mimic the board-level counterfeiting situation by referring any Arduino Uno boards as ``counterfeit'' boards with the exception that official Arduino Uno boards are treated as the genuine ones. 
% and the boards fabricated by other vendors as the counterfeit ones.
We purchase a total of 39 boards from three different vendors (13 from each vendor): Arduino.cc, Elegoo, and Kuman, as shown in Fig.~\ref{fig:arduino_uno_layout}(a)-(c), respectively.
All three designs share the same schematic and layout.
We mimic PCB cloning by referring to Elegoo and Kuman boards as counterfeit boards, while the official Arduino boards from Arduino.cc are treated as genuine ones. 

% The recycling detection is performed on the official Arduino Uno board by stressing 4 boards using a temperature chamber with $100^\circ\text{C}$ ($373\text{K}$) to mimic worn-out boards. 
% The according thermal acceleration factor (TAF)~\cite{maes2012experimental} is 61.05,\footnote{$TAF=\exp\bigl(\frac{E_a}{k}(\frac{1}{T_{op}}-\frac{1}{T_{stress}})\bigr)$, where $k=8.62\times10^{-5}eV/\text{K}$, $E_a=0.5eV$,  $T_{op}=295\text{K}$ is nominal temperature, $T_{stress}=373\text{K}$ is stressed temperature.}
% and the boards are stressed for 71 hours to have roughly 6 months of equivalent use time.

\smallskip
\noindent\textbf{Detection Results Analysis.}
% \subsection{Arduino Uno from Different Vendors}
Fig.~\ref{fig:arduino_uno_layout}(d) illustrates \DesName{} can successfully detect the cloned boards, where we show $Z_{33}$ out of 6 collected $Z$-parameters.
During the measurements, we notice that there are two batches of Elegoo boards (marked as Elegoo-I and Elegoo-II), which cannot be visually distinguished from each other.
However, we can confirm the two batches from their impedance profiles.
The impedance profiles of Elegoo-I and Elegoo-II are different from each other for all $Z$-parameters. For each batch, the impedance profiles of the boards are consistent with other boards of the same batch.
Since the differences are mostly at the low frequency, we infer that the two batches have the same PCB layout, but the on-board components may be from different vendors or have different specifications.

\begin{comment}
\begin{figure}[t]
\centering
\includegraphics[width=3.3in]{figs/arduino_layout_beta.pdf}
\caption{(a) The layout of Arduino Uno board design with three measurement ports highlighted. (b)-(d) The photographs of Arduino Uno boards from three different vendors, respectively: official Arduino, Elegoo, and Kuman.}
\label{fig:arduino_uno_layout}
\end{figure}
\end{comment}

% \textit{Cloning:}
In Fig.~\ref{fig:arduino_uno_layout}(e), we plot the $FD'$ histogram of the three vendors with respect to official Arduino boards, to show that the genuine Arduino boards can be distinguished from the counterfeit (i.e., non-Arduino) boards with $100\%$ TPR and $0\%$ FPR.
We also demonstrate that the developed $FD'$-KNN algorithm can classify the boards into multiple classes to prevent false positives on different batches of boards.
Moreover, multi-port measurement can increase the classification accuracy.
Fig.~\ref{fig:arduino_uno_layout}(f) illustrates the relationship between the average classification accuracy, the number of boards for training, and the number of ports.
We use $FD'$-3NN to classify the board into 4 classes (i.e., Arduino, Kuman, Elegoo-I, and Elegoo-II). 
For each configuration (i.e., $\#$ of training boards and $\#$ of ports), we run 500 trials and compute the average accuracy.
Both the training boards and the measured ports are randomly selected for each trial.
Compared with 1-port detection, 2-port detection improves the classification accuracy to higher than 99.8\%.
% There are errors in classification for the reason that the impedance profiles of Kuman boards are similar to the Elegoo-I.
In addition, using the 3-port detection, a 100\% detection accuracy can be reliably achieved with 4 training boards from each class.

% % % 
% % \noindent\textbf{Recycled Boards Detection.} 
% \textit{Recycling:}
% In Fig.~\ref{fig:arduino_uno_layout}(g) we compare the impedance profiles of both reference and recycled boards.
% % We find a deviation in the impedance profile due to recycling.
% The recycled boards have lower impedance at $1.2MHz$, which is mainly due to the increasing leakage current of the capacitors. 
% We observe 34.3 average inter-$FD'$ with respect to genuine Arduino boards, corresponding to up to $62.5\%$ TPR and $33\%$ FPR. Note that this performance is mainly limited by the short amount of time in our experiments to stress the board.
% In practice, recycling is performed beyond 6-month use time to reap the maximal profits.
% % thus causing the lower impedance at the dump.
% % Since the recycled boards have lower impedance, we believe this mainly due to increasing leakage current of the capacitors, thus causing the lower impedance at the dump.
% \DesName{} can detect recycling boards effectively. 
% % The results show that we are able to detect the recycling.

% % \begin{figure}[t]
% % \centering
% % \includegraphics[width=3.3in]{figs/arduino-uno-recycled.pdf}
% % \caption{P22 \huifeng{will combine this figure and figure~\ref{fig:arduino_uno_results}}}
% % \label{fig:arduino_uno_recycled_results}
% % \end{figure}

%% file: doc/discussion.tex
\label{sec:discussion}

% \noindent\textbf{Detection Sensitivity.}

{In this section, we discuss \DesName{}'s capabilities against adaptive attackers who attempt to bypass \DesName{} intentionally.
The detection sensitivity is first explored to show the performance of detecting well-designed Trojans. Then other stealthier mechanisms that attackers can utilize are discussed.
Overall, \DesName{} aims to create an effective security asymmetry between attack and defense.
Although \DesName{} is not an ultimate solution for PCB attacks, implementing \DesName{} can significantly mitigate potential threats. The most motivated attackers can intentionally bypass \DesName{} but likely at the cost of making their malicious implants more easily detected by orthogonal approaches (e.g., inspection, functional test, and integrity check), or requiring significantly advanced techniques.}

\subsection{Detection Sensitivity}\label{sec:sensitivity}

To achieve both visual and electrical stealth, attackers may deliberately miniaturize malicious circuits (e.g., by using small-package chips). This strategy aims to bypass \DesName{} anomaly detection because chips/components with small footprints also tend to have lower parasitics, making them closer to the ideal open circuit ($C$=0, $L$=$\infty$, $R$=$\infty$) when connected in parallel with the PDN.   
However, in frequency-domain PDN analysis, malicious modifications with minimal parasitics are still observable as shifted PDN profiles with discernible magnitudes~\cite{kim2010chip} at high frequencies. 
% (see Fig.~\ref{fig:pds_methodology}).
This property facilitates \DesName{}'s detection sensitivity. 

\begin{figure}[t]
\centering
\includegraphics[width=3.45in]{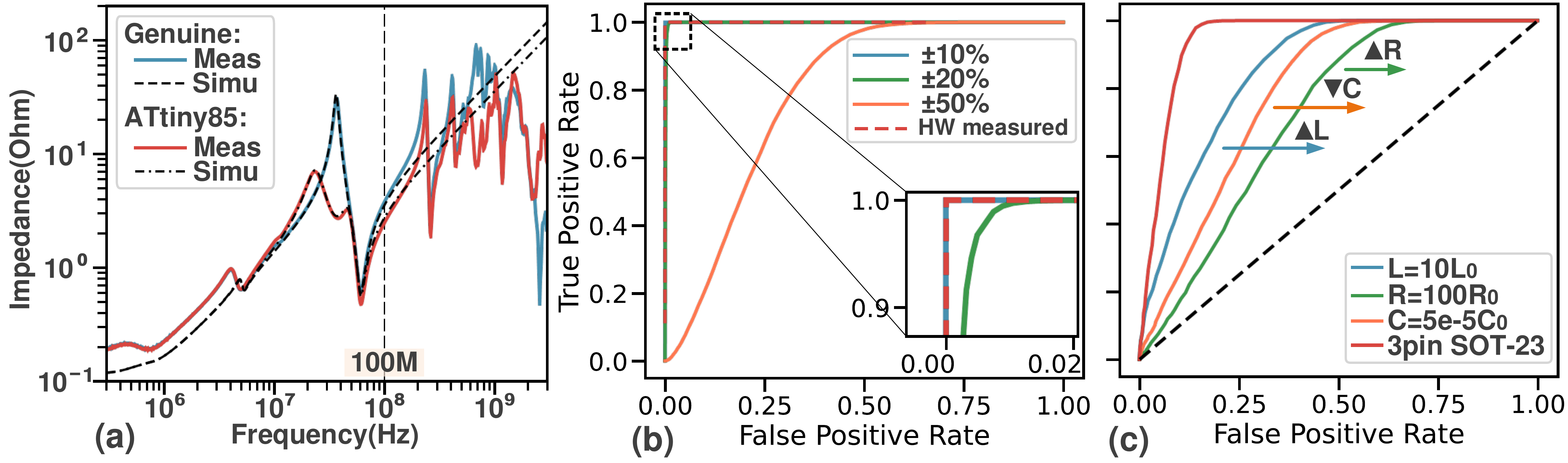}

\caption{(a) Measured and simulated PDN profiles of customized PCB. Simulated ROC curves (b) with different PCB tolerances and (c) when taking ATtiny85 ($C_0$,$L_0$,$R_0$) as a reference and scaling the malicious modifications. The results show \DesName{}'s sensitivity.}
% \vspace{-0.1in}
\label{fig:pdnpulse1v3_results_simu}
\end{figure}

To investigate the limits of detection sensitivity of \DesName{} in response to attackers' efforts, we build a simulation model to capture the PDN of the custom PoC boards in Sec.\ref{sec:pdnpulse1v3} and study the impact of various malicious modifications. We use the well-accepted modeling methodology described in the industry documentation~\cite{intelpdn} and abstract the chips as RLC networks in parallel with the PDN. The circuit model is validated by toggling the connection/disconnection of ATtiny85 to the PDN, then comparing the simulated impedance profiles with the experimentally measured one. The simulation results are consistent with the hardware measurements (see the black dashed lines in Fig.~\ref{fig:pdnpulse1v3_results_simu}(a)). Note that we focus on modeling the parasitics of custom PCB and anomalies. The analysis of board resonance (e.g., $>$$100MHz$ for custom PCB) is out of the scope of this work. 

The sensitivity of \DesName{} is affected by both PCB process variation (i.e., tolerance) and intrinsic parasitics of the anomalies.
In Fig.~\ref{fig:pdnpulse1v3_results_simu}(b), we present the simulated \DesName{} performance for detecting ATtiny85 with varying levels of PCB tolerances using 
Monte-Carlo simulation, where process variation follows Gaussian distribution and the tolerance is $3\sigma$. Results in Fig.~\ref{fig:pdnpulse1v3_results_simu}(b) show that \DesName{} has desirable performance under $\pm20\%$ tolerance, beyond the worst case variations for COTS boards. Acceptable performance can be achieved even with $\pm50\%$ tolerance. We thus conclude that \DesName{} detection sensitivity is sufficient to handle process variation of typical COTS boards. This conclusion is further validated by comparing the simulated ROC curves with hardware measurements (the red dashed line in Fig.~\ref{fig:pdnpulse1v3_results_simu}(b)).

% \xuan{this paragraph needs revising.} 
In Fig.~\ref{fig:pdnpulse1v3_results_simu}(c), we present \DesName{}'s sensitivity as a function of the RLC parameters. %associated with modifications of small physical dimensions.  
%In this experiment, the frequency of interest is increased to $300KHz$$\sim$$4GHz$.
%The results are also based on Monte-Carlo simulation with $\pm10\%$ tolerance. 
% The parasitics of scaled chip are $(C\times k, L\times k, R/k)$, where  are the parasitics of ATtiny85.
% In a similar manner, we also simulate 
We first simulate the chip with 3pin SOT-23 package ($C$=0.12$pF$,$L$=1.4$nH$,$R$=3$\Omega$)~\cite{ma4e1340_sot23},
one of the smallest package footprints available
% ($\times0.8$ smaller than the tiniest COTS package for MCU as claimed in~\cite{smallest_microcontroller}).
(2.6mm$\times$2.9mm).
Results show \DesName{} can successfully spot these stealthy changes. 
We then use the parasitics of an ATtiny85 chip ($C_0$=0.9$nF$,$L_0$=21$nH$,$R_0$=3$\Omega$) as a reference and scale its RLC values to explore the limits of \DesName{}.
% We present the trend of detecting performance changes with respect to each parameter. 
We can achieve acceptable performance with even $10L_0$, $100R_0$, or $5$$\times$$10^{-5}C_0$($1.8fF$), where the parasitics of anomaly are an order-of-magnitude smaller than the SOT-23 package.
Even though the simulation is based on the Trojan PoC board, we are confident that the conclusion and trends here can be extended to other types of anomalies including counterfeits.
Note that the RLC network model is not specific to ATtiny85 and can represent any type of anomaly described in this paper.
By including the board resonance, adopting multi-port detection, and increasing the measurement bandwidth, the performance can be further improved.

\begin{figure}[t]
\centering
\includegraphics[width=\columnwidth]{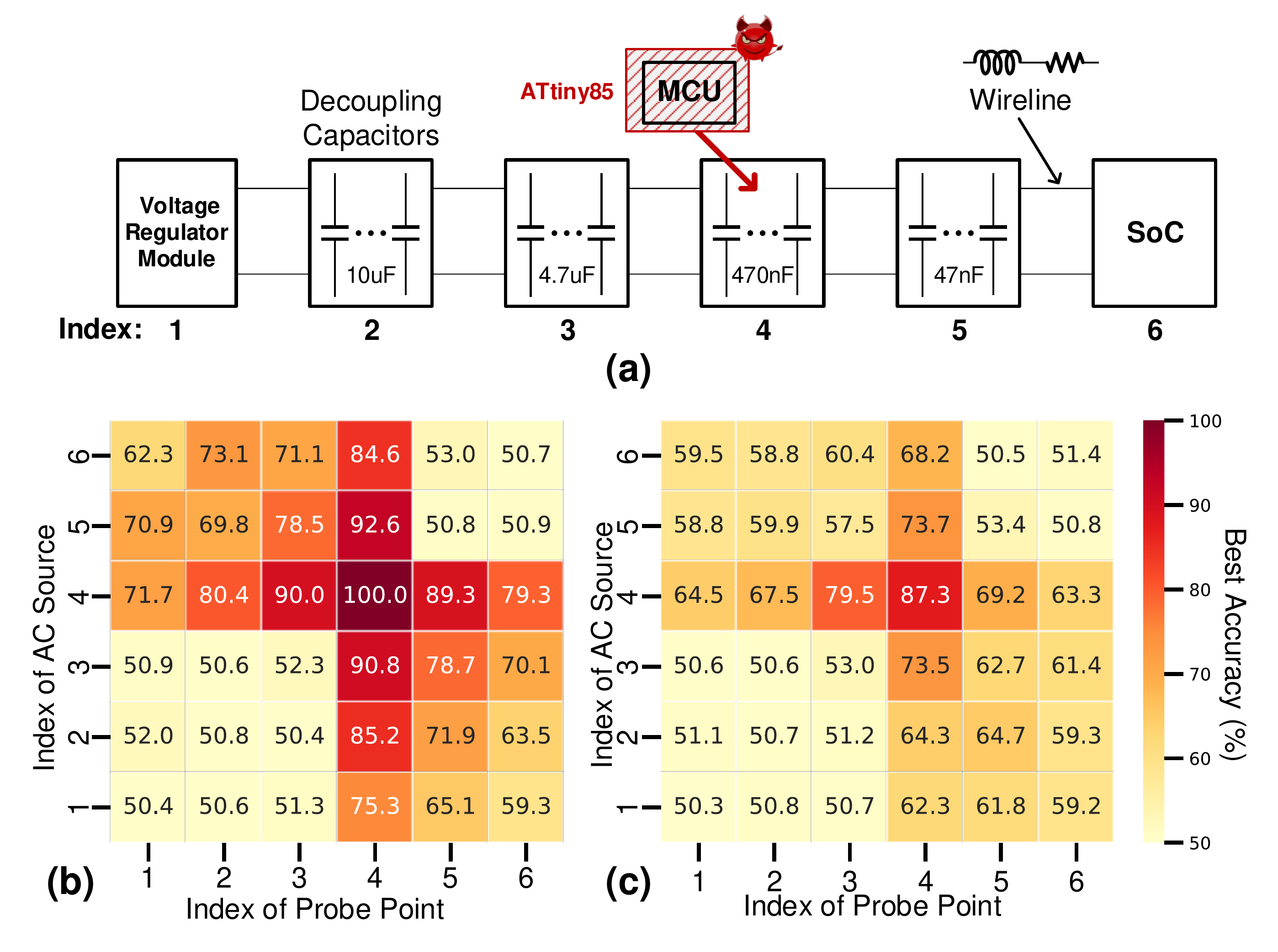}
\caption{(a) The simplified structure of updated PDN structure in simulation. The best anomaly detection accuracy with varying locations of AC source and probe point when PDN has a total of (b) 6 capacitors and (c) 12 capacitors.}
% \vspace{-0.1in}
\label{fig:decap_role_simu}
\end{figure}

Besides miniaturizing malicious circuits, attackers may exploit capacitors' isolation effect to bypass PDNPulse. Precisely, attackers can place the anomalies in close proximity to the decoupling capacitors, making the anomalies connected in parallel with low-impedance capacitors. We conduct a simulation by updating the previous PDN model to study the potential for attackers to conceal anomalies. 
The updated PDN structure is shown in Fig.~\ref{fig:decap_role_simu} (a) and includes four groups of decoupling capacitors with values of 10uF, 4.7uF, 470nF, and 47nF, whose models are adopted from~\cite{intelpdn}. 
The four groups contain 1, 1, 2, and 2 capacitors, respectively.
We also include a wireline model~\cite{zhu2020powerscout} to connect the VRM, the capacitor groups, and the SoC.
For clarity, we index the locations of VRM, each group of capacitors, and SoC from 1 to 6.
An ATtiny85 chip, mimicking anomaly, is inserted in the 470nF capacitor group (i.e., index 4) to attempt to bypass PDNPulse.
Similar to the previous simulation, we use the Monte-Carlo simulation and set the process variation as $\pm$10\% tolerance.

Fig.~\ref{fig:decap_role_simu} (b) illustrate the best anomaly detection accuracy (i.e., $(TP+TN)/(TP+TN+FP+FN)$).
By setting the AC source $x$ and probe point $y$ to different indexes, we measure various transfer impedance (e.g., $Z_{21}$ if AC source index $x$=1 and probe point index $y$=2). 
Note that when $x$=$y$, the measured impedance is self-impedance. 
The results indicate that if the detection is only based on the self-impedance (i.e., $x$=$y$) unless the measurement point is at the same index as the ATtiny85 chip, we fail effectively detect the chip (with maximum 52.3\% accuracy).
However, leveraging the transfer impedance, the ATtiny85 chip can be acceptably detected as long as the measured transfer impedance covers the location of the ATtiny85 chip (e.g., $x$$>$4 and $y$$<$4).
We also find that the detection accuracy decrease as the increase of the distance between AC source and probe point ($|$$x$-$y$$|$).
For example, with $x$=$1$ and $y$=$6$, the accuracy decreases to only 59.3\%. 

We further investigate the impact of implementing more capacitors in the PDN.
The number of capacitors in each group is doubled, and the results are presented in Fig.~\ref{fig:decap_role_simu} (c).
Compared to the previous configuration, the accuracy decreases for both self-impedance and transfer impedance.
% This 
The decline in accuracy results from more capacitors causing a lower impedance, making anomaly impacts less obvious.
Please note that here we consider only using at most two ports to detect anomalies.
The accuracy can be increased by measuring from more ports and applying KNN algorithms.

\begin{table}[t]
\footnotesize
\centering
\caption{Comparison of detection accuracy on Arduino Due boards using single port vs. four ports under various port selection scenarios.}
\label{tab:multi_port_tradeoff}
\begin{tabular}{@{}cccccc@{}}
\toprule
Single Port & 84.4\% & 73.3\% & 62.2\% & 91.1\% & 91.1\% \\ \midrule
Four Ports  & 100\%  & 100\%  & 100\%  & 100\%  & 100\%  \\ \bottomrule
\end{tabular}
\end{table}

\subsection{Multi-Port Detection Trade-offs}

This subsection examines the advantages of multi-port detection in comparison to single-port detection using experiments on Arduino Due boards. It contrasts the detection accuracy when using a single port versus employing four ports under various port selections. The results are listed in Table~\ref{tab:multi_port_tradeoff}. For the Arduino Due board, where the PDN is abstracted into a five-port network, we iteratively select single port/four-port combinations to conduct detection. By using multi-port detection, PDNPulse can achieve a stable full coverage of all anomalies, which cannot be achieved by single-port detection.

Multi-port detection entails certain costs, which can be assessed from two perspectives: the number of measurements (represented by $(n^2 + n)/2$ for $n$ port detection) and the supplementary computational expenditures linked to calculating the S- to Z-parameter conversion (as shown in Equation~(\ref{eq:s2z_matrix})) and the multi-port $FD'$ (as illustrated in Equation~(\ref{eq:fd_multi_ports})). Nonetheless, PDNPulse is tailored for static detection, allowing these overheads to be effectively amortized through parallelization with other standard PCB tests. Hence, the advantages provided by multi-port detection significantly outweigh the incremental costs for security-critical applications.

\subsection{Attacker Response}\label{sec:attacker_response}
{
% PDNPulse focuses on physical modifications of the PCB that exert persistent effects in the electrical domain. 
% Our work is complementary to the existing works and has many additional advantages (will be discussed in Sec~\ref{sec:relatedwork}).
% Experimental results have shown that \DesName{} has a full coverage on mitigating counterfeit chips/components/PCBs.
% While for other attacks, 
We explore available attack vectors that either undermine \DesName{} by avoiding connection to the PDN, or attempt to bypass \DesName{} by hiding its impact on the PDN impedance profile and discuss their feasibility in what follows.

% We explore available attack vectors and discuss their feasibility.

% Below we discuss the feasibility of potential mechanisms that can be leveraged by attackers:

% \begin{itemize}[leftmargin=*,topsep=0px,partopsep=0px]
    % \item \textbf{Avoid Connecting to PDN.}
    
    \medskip
    \subsubsection{\textbf{Avoid PDN Connection}}
    
    \DesName{} assumes that counterfeits and persistent PCB Trojans require a power supply to operate and that their power pins are typically attached to the PDN. Thus, adversaries can avoid connecting the PDN to bypass \DesName{}.
    
    \smallskip
    % \begin{description}
    \noindent\textbf{{Using an Alternative Power Source.}}
    % To avoid attaching to PDN but still can obtain the power supply, a
    An adversary may use self-powered circuits or harvest power from data signals to avoid connecting directly to the PDN.
    %Self-powering includes equipping with a battery or harvesting ambient energy.
    If a self-powered Trojan employs a battery that is intended to be both persistent and long-term, then both its capacity and size need to meet those specifications (e.g., to spy on the target system~\cite{robertson2018big}). COTS batteries are not an option in this case since the diameters of most are greater than 6mm~\cite{battery_size}. 
    % which can be even larger than the Trojan itself. 
    Thus, attackers would need to adopt significantly advanced battery fabrication and integration techniques to be successful. 
    Circuits that harvest ambient energy are typically based on coils or photodiodes with relatively large footprints to maximize energy extraction. In this case, the main challenge of the adversary would be evading visual inspection. 
    % If this can be addressed, then a successful attack can be launched against \DesName{}. 
    However, we expect multiple anomaly detection techniques to be employed, such as ones that detect hidden modifications~\cite{asadizanjani2015non,iqbal2017pcb}, because \DesName{} is not intended to be an ultimate solution. 
    %Thus, attackers should worry about how to evade visual inspection instead. 
    %Besides, the reliability will be a concern since attackers cannot control the amount of harvested ambient energy once Trojans are implemented, causing attackers may not have a reliable control on the Trojans to perform the attack.  
    % All the above evade methods could be easily detected by visual inspection.
    % Although 
    Trojans that harvest power from data signals are also plausible. Note that a practical implementation should consider the loading effect on the signal pins and avoid impacting the signal quality to pass the standard functionality tests. Another consideration for such an attack is that to provide a stable power supply the adversary would also need to control the status (e.g., keep logic high) of that signal pin when the Trojan is working, which requires additional handling circuitry.
    It is worth mentioning that attacks using the above three methods have not been reported. 
    % in existing works~\cite{harrison2021malicious,zhu2021pcbench,mehta2020big}. \dean{Can we refer to these works as SoK's or use language that clarifies that they in part survey existing board-level attacks?}\huifeng{Maybe just delete "in existing works[]"?}
    % and thus be detectable by functionality testing. 
    
    \smallskip
    \noindent\textbf{{Passive Attacks.}} Attackers may also use purely passive circuits which do not require a power supply. However, we believe the potential attacks using only passive circuits are limited (typically to signal traces) and often lead to uncontrollably compromise of essential functions.
    For example, an attack is presented in~\cite{ghosh2014secure} that alters trace spacing and dimensions to cause cross-talk, but also impacts the original signal quality due to increased interference. 
    Detecting such modifications to signal traces is out of the scope of our work but it can be achieved by functionality tests.
    Please note that \DesName{} can detect passive attacks related to the PDN (see Sec.\ref{sec:pdnpulse1v3}).
    Designing configurable and stealthy Trojans using passive circuits is still an open question, and we are unaware of any documented attacks. 
    
    \medskip
    % \item \textbf{Constrain Impact on PDN Profile.}
    \subsubsection{\textbf{Hidden Impact on PDN Impedance Profile}}
    Although Trojans have to attach to the PDN, attackers may try to reduce their impact on PDN profiles to evade \DesName{}. We discuss several possible methods for doing so below.

    \smallskip
    \noindent\textbf{Conceal in the Decoupling Capacitors.}
    Adversaries mayhide anomalies near decoupling capacitors to evade PDNPulse detection. But this strategy creates a challenge in maintaining short wiring distance from the anomalies to the desired point on the PDN while also keeping the wiring distance from the anomalies to the payload short. If this challenge is not addressed, the anomalies may not be stealthy enough. Our earlier discussion (Sec.~\ref{sec:sensitivity}) has shown that the proposed transfer impedance-based detection methods can effectively detect anomalies hiding in capacitors. However, we also found that for large-scale and high-performance PCBs that require ultra-low impedance PDN, detection becomes more challenging as more capacitors are implemented. One way to improve detection performance is to increase the number of measurement ports and include board resonance in the analysis.

    \smallskip
    % \begin{description}
    \noindent\textbf{{Exploit Insufficient Measurements.}} Adversaries may learn the measurement parameters of \DesName{}, such as the measured voltage domains and ports, and then intentionally implant their Trojan outside of those parameters to avoid being detected. 
    The challenge for this approach is that malicious modification will unpredictably affect multiple impedance profiles (i.e., multiple $Z$-parameters) simultaneously (see Sec.~\ref{sec:pdn_sensitivity}).
    To find reliable measurement blind spots for implantation under multi-domain multi-port detection, attackers would need to obtain the RF models of both the victim PCB and Trojan circuits, then simulate the PDN profiles when placing Trojans in each potential location to verify the blind spot will remain hidden.
    Note that such RF models are usually not available and attackers typically need to measure the whole PDN using our method to obtain the models. 

    % One practical strategy can be making malicious implants affect PDN profile higher than measured spectrum bandwidth (e.g., $3GHz$ in this work). 
    % This strategy would significantly increase difficulties in controlling the Trojan size and its parasitic effects, and can be thwarted by using VNA with higher specifications

    \smallskip
    \noindent\textbf{{Compensate for PDN Impedance Effect.}} As described in the previous section, a motivated attacker may obtain the RF model of the PDN. They may then adjust the circuit design to compensate for the parasitics of the Trojan and avoid detection. 
    Unfortunately, passive R, L, and C components have different (instead of mutually offset) effects in the frequency domain, which prevents them from canceling one another out. However, they can shift the affected spectrum band (i.e., compensate the impedance at one or several frequency points).
    To fully compensate for the PDN profiles, attackers would have to adjust the original PDN design (e.g., remove decoupling capacitors). 
    However, since the adjustment for one $Z$-parameter will inevitably affect other $Z$-parameters attackers must exhaustively search for a solution to fully compensate the PDN profile (all $Z$-parameters), and there may even be no such solution.
    
    \smallskip
    \noindent\textbf{{Transient Physical Modifications.}} \DesName{} is not designed as a tamper-evident technology and attacks that can be undone, or are not persistent physical modifications, are outside its scope. One possible attack can be de-soldering, maliciously programming, then re-soldering the memory chip~\cite{exploitee2017all}. However, such attack can be caught by software integrity verification~\cite{suh2003efficient}.
    % \end{description}
% \end{itemize}
}

%% file: doc/future_work.tex
% \subsection{Limitations and Caveats}
\label{sec:limitation}

Please note that \DesName{} could suffer detection failures when insufficient or inappropriate ports/voltage domains are measured.
% By properly selecting voltage domains and ports, \DesName{} is shown to comprehensively detect PCB anomalies. 
We have analyzed its robustness to port selection (e.g., Fig.~\ref{fig:arduino_due} and~\ref{fig:arduino_uno_layout}) in multi-port detection.  
However, the strategy for determining the minimum number of ports and the most appropriate port/voltage domain is device-specific and anomaly-specific. 
% we did not explore the minimum number and the most efficient selection strategy of port/voltage domain selection since they are device-specific and anomaly-specific. 
We plan to investigate it with a complex distributed simulation model in future work.

In practical conditions, the tolerance of the golden model can be affected by various factors such as production process and manufacturing defects, making it challenging to determine the optimal deviation value.  
One limitation of our proposed method, PDNPulse, is that it can be difficult to distinguish between manufacturing flaws and malicious modifications in a PCB, as both can result in anomalies. 
For example, as shown in Fig.~\ref{fig:arduino_due}(c), the impedance profile of the M3 board is surprisingly different from the others, but we could not see any difference between the M3 and the other four malicious boards. 
% This unusual situation can also occur with genuine boards and raise a false alarm.
That is, confusion about manufacturing flaws and malicious modifications may lead to a false security alarm.
Interpreting the PDN impedance profile can solve this issue and will be a focus of our future work.

We have shown that \DesName{} is robust to non-anomaly changes such as de-soldering and replacing with the same IC (e.g., Fig.~\ref{fig:motherboard_layout}(e)) as well as typical process variation (e.g., Fig.~\ref{fig:pdnpulse1v3_results_simu} (b)). However, to prevent false positives, the process variation due to different batches/vendors should be included in building the golden model. For instance, in Fig.~\ref{fig:pynq_results}(f), due to the large process variation between chip batches, directly applying the $FD'$ algorithm will yield false positives. Our $FD'$-KNN method can effectively deal with such situations by classifying the board under test into multiple vendors/batches, as shown in Sec.\ref{sec:arduino_uno}.

% In this paper, although the detected Trojans include inner-layer changes, we did not explicitly prove \DesName{} capability for detecting mere inner-layer modifications. However, \DesName{} leverages monitoring board-level PDN impedance profiles. These are  electrical properties and thus intrinsically let \DesName{} not be limited by the layers of modifications. We will leave the validation of such capabilities to our future work.

% \smallskip
% \noindent\textbf{Attacker Response.}

{
One of the salient features of \DesName{} is it has no hardware overhead and can be applied to legacy PCBs without any modifications. Potential PCB changes such as adding test points connected to the PDN can be made to increase \DesName{}'s stability and performance. A \DesName{}-aware PCB design framework will be our future direction, which can efficiently insert PDN test points during the design stage.
}

%% file: doc/related_works.tex
Facing numerous board-level attacks, detection methods have been developed, but they are often piecemeal solutions, capable of identifying only a particular attack under specific restrictions. 
% Table~\ref{tab:my-table} highlights that \DesName{} can rapidly capture a wide range of threats at relatively low cost.

\smallskip
\noindent\textbf{Reverse Engineering and Image Inspections.} 
Reverse engineering~\cite{grand2014printed, botero2020hardware} provide most comprehensive detection, but it suffers from long detection time and high cost, and it is destructive.  
Image inspection methods can be divided into surface imaging and volumetric imaging. 
Surface imaging uses such as visible light~\cite{xie2013detecting} and interferometry~\cite{mehta2020big} to detect anomalies. 
Cameras or microscopes are needed to detect the change of PCB surface pattern~\cite{iqbal2017pcb} or to visually examine PCBs~\cite{de2017detecting}. 
However, surface imaging cannot detect sophisticated counterfeits.
For volumetric imaging using radiation, X-ray~\cite{asadizanjani2015non,iqbal2017pcb} is commonly applied to comprehensively capture the internal structure of PCBs, which also suffers high cost and needs to de-solder components.
% \footnote{X-ray typically requires de-soldering the surface chips/components.} 
In contrast, our method does not require expensive optical/X-ray equipment and can be conducted with a standard VNA.

\smallskip
\noindent\textbf{Side-Channel Analysis.} System-level delay side-channel information
(such as based on JTAG~\cite{hennessy2016jtag}, I2C~\cite{shwartz2020inner}),
power side-channel information~\cite{piliposyan2020hardware}, or combined
multiparameter side-channel analysis~\cite{ghosh2014secure} can be utilized to
perform anomaly detection and run-time monitoring. 
% For both power and
% delay side-channel information, they are highly related to the PDN of the
% system. 
Although, power and delay side-channel leakage are fundamentally caused by the PDN of target devices, 
% Power side-channel leakage is due to the convolution between current draw and PDN impedance. 
% While, the induced supply voltage fluctuations will impact logical components and further cause delay side-channel leakage.
% PCB side-channel 
their analysis is usually limited to specific anomaly types and offers only partial PCB area coverage.
Rather than relying on leaked information, we directly measure the PDN and thus can retrieve more in-depth
information. 

\smallskip
\noindent\textbf{Impedance Measurement.} 
The changes to impedance patterns, resonant frequency, signal response of the trace, bus, or transmission line can be measured to detect anomalies~\cite{mcguire2019pcb,zhang2015robust,guo2017mpa,wei2019transmission,paley2016active,xu2020bus,fujimoto2018demonstration}. 
However, due to using parts of the PCB design (e.g., bus), these approaches can only detect the anomalies attached to this trace, limiting both detectable anomaly types and locations. 
Since \DesName{} is based on the PDN, which is connected to each part of the system, the coverage of detection is significantly increased.

\smallskip
\noindent\textbf{PDN Impedance-based Measurement.} 
{PDN impedance-based detection has been explored in various ways~\cite{nishizawa2018capacitance,wang2019system}. 
As discussed in Sec.\ref{sec:intro}, although successful, existing PDN impedance solutions only focus on specific types of anomalies and parts features (e.g., impedance at fixed frequencies) of the local PDN.
Our work comprehensively explores the characteristics of PDN of the whole PCB. 
We specify the systematic analysis of PDN effects and the experimental setup for accurate multi-port, multi-domain PDN measurements and demonstrate experimentally PDNPulse’s robustness to probe location, PCB scale, and port numbers.
Our extensive experimental results show \DesName{}'s effectiveness in detecting board-level attacks and counterfeiting.}

In recent work~\cite{mosavirik2022scatterverif, mosavirik2022impedanceverif}, the authors proposed a PCB tampering and counterfeit detection framework based on monitoring changes in the scatter parameters (i.e., S-parameters) of the PDN. Our work, PDNPulse, is distinct from this approach as we utilize multi-port and multi-domain detection, measuring transfer impedance between ports to detect system-level anomalies. Additionally, we use Z-parameters, which are commonly used in modeling PDN~\cite{gupta2007understanding,swaminathan_power_2004,ren_frequency-dependent_2009,kim_system_2008}, allowing for direct comparison with simulation results to better interpret and model the impacts of anomalies. Our measurement setup also utilizes a customized probe that can be applied to legacy systems and attached to any point on the PCB, eliminating the need for hardware modifications and allowing for multi-port detection. Our experiments cover a wide range of PCBs, demonstrating the robustness and effectiveness of our proposed method in various design scenarios.
Overall, our work achieves a unique contribution to the field.

%% file: doc/conclusion.tex
We propose a novel board-level attack detection framework named \DesName{}.
It leverages the inherent sensitivity of the on-board PDN to reliably authenticate that a PCB is free from tampering and/or anomalies.
It is light-weight and compatible with legacy systems, and requires no hardware overheads or design modifications for deployment.
We conduct extensive experiments on custom and COTS PCBs covering different design scales, anomaly types, and threat models. We demonstrate that \DesName{} can capture a wide range of threats at a low cost.

%% file: TIFS_PDNPulse.bbl
\begin{thebibliography}{10}

\bibitem{exploitee2017all}
All your things are belong to us.
\newblock \url{https://www.exploitee.rs/}, 2017.
\newblock DEF CON.

\bibitem{arduino_due}
Arduino due.
\newblock \url{https://store.arduino.cc/usa/due}, 2020.

\bibitem{beaglebone_wireless}
Beaglebone-wireless.
\newblock \url{https://github.com/beagleboard/beaglebone-black-wireless}, 2020.

\bibitem{seaborn_pi}
Statistical estimation and error bars - seaborn.
\newblock
  \url{https://seaborn.pydata.org/tutorial/error_bars.html#percentile-interval-error-bars},
  2023.

\bibitem{wiki_parallel}
Wikipedia: Series and parallel circuits.
\newblock \url{https://en.wikipedia.org/wiki/Series_and_parallel_circuits},
  2023.

\bibitem{ArduinoUno}
Arduino.CC.
\newblock Arduino uno rev3.
\newblock \url{https://store.arduino.cc/usa/arduino-uno-rev3}, 2021.
\newblock Accessed April, 2021.

\bibitem{asadizanjani2015non}
Navid Asadizanjani, Sina Shahbazmohamadi, Mark Tehranipoor, and Domenic Forte.
\newblock Non-destructive pcb reverse engineering using x-ray micro computed
  tomography.
\newblock In {\em 41st International symposium for testing and failure
  analysis, ASM}, pages 1--5, 2015.

\bibitem{awal2022nearfield}
Md~Sadik Awal, Arjuna Madanayake, and Md~Tauhidur Rahman.
\newblock Nearfield rf sensing for feature-detection and algorithmic
  classification of tamper attacks.
\newblock {\em IEEE Journal of Radio Frequency Identification}, 6:490--499,
  2022.

\bibitem{bhunia2018hardware}
Swarup Bhunia and Mark Tehranipoor.
\newblock {\em Hardware security: a hands-on learning approach}.
\newblock Morgan Kaufmann, 2018.

\bibitem{bogatin2010signal}
Eric Bogatin.
\newblock {\em Signal and power integrity--simplified}.
\newblock Pearson Education, 2010.

\bibitem{botero2020hardware}
Ulbert~J Botero, Ronald Wilson, Hangwei Lu, Mir~Tanjidur Rahman, Mukhil~A
  Mallaiyan, Fatemeh Ganji, Navid Asadizanjani, Mark~M Tehranipoor, Damon~L
  Woodard, and Domenic Forte.
\newblock Hardware trust and assurance through reverse engineering: A survey
  and outlook from image analysis and machine learning perspectives.
\newblock {\em arXiv preprint arXiv:2002.04210}, 2020.

\bibitem{SOLT}
NATIONAL~INSTRUMENTS CORP.
\newblock Short-open-load-through (solt) calibration.
\newblock
  \url{https://www.ni.com/docs/en-US/bundle/ni-vna/page/vnahelp/calibration_solt.html},
  2023.

\bibitem{de2017detecting}
Thomas Jose~Mazon De~Oliveira and et.al.
\newblock Detecting modifications in printed circuit boards from fuel pump
  controllers.
\newblock In {\em 30th SIBGRAPI Conference on Graphics, Patterns and Images
  (SIBGRAPI)}. IEEE, 2017.

\bibitem{eiter1994computing}
Thomas Eiter and Heikki Mannila.
\newblock Computing discrete fr{\'e}chet distance.
\newblock Technical report, Citeseer, 1994.

\bibitem{flynn2018power}
Colin Flynn.
\newblock I, for one, welcome our new power analysis overlords.
\newblock
  \url{https://i.blackhat.com/us-18/Wed-August-8/us-18-OFlynn-I-For-One-Welcome-Our-New-Power-Analysis-Overloards.pdf},
  2018.
\newblock Black Hat USA 2018.

\bibitem{fujimoto2018demonstration}
Daisuke Fujimoto, Shota Nin, Yu-Ichi Hayashi, Noriyuki Miura, Makoto Nagata,
  and Tsutomu Matsumoto.
\newblock A demonstration of a ht-detection method based on impedance
  measurements of the wiring around ics.
\newblock {\em IEEE Transactions on Circuits and Systems II: Express Briefs},
  65(10):1320--1324, 2018.

\bibitem{ghosh2014secure}
Swaroop Ghosh, Abhishek Basak, and Swarup Bhunia.
\newblock How secure are printed circuit boards against trojan attacks?
\newblock {\em IEEE Design \& Test}, 32(2), 2014.

\bibitem{grand2014printed}
Joe Grand.
\newblock Printed circuit board deconstruction techniques.
\newblock In {\em 8th $\{$USENIX$\}$ Workshop on Offensive Technologies
  ($\{$WOOT$\}$ 14)}, 2014.

\bibitem{greenberg2019tinyspy}
Andy Greenberg.
\newblock A new proof-of-concept hardware implant shows how easy it may be to
  hide malicious chips inside it equipment.
\newblock
  \url{https://www.wired.com/story/plant-spy-chips-hardware-supermicro-cheap-proof-of-concept/},
  2019.

\bibitem{guin2014counterfeit}
Ujjwal Guin, Ke~Huang, Daniel DiMase, John~M Carulli, Mohammad Tehranipoor, and
  Yiorgos Makris.
\newblock Counterfeit integrated circuits: A rising threat in the global
  semiconductor supply chain.
\newblock {\em Proceedings of the IEEE}, 102(8):1207--1228, 2014.

\bibitem{guo2017mpa}
Zimu Guo, Xiaolin Xu, Mark~M Tehranipoor, and Domenic Forte.
\newblock Mpa: Model-assisted pcb attestation via board-level ro and
  temperature compensation.
\newblock In {\em 2017 Asian Hardware Oriented Security and Trust Symposium
  (AsianHOST)}, pages 25--30. IEEE, 2017.

\bibitem{gupta2007understanding}
Meeta~S Gupta, Jarod~L Oatley, Russ Joseph, Gu-Yeon Wei, and David~M Brooks.
\newblock Understanding voltage variations in chip multiprocessors using a
  distributed power-delivery network.
\newblock In {\em Proceedings of the conference on Design, automation and test
  in Europe}, pages 624--629, 2007.

\bibitem{harrison2021malicious}
Jacob Harrison, Navid Asadizanjani, and Mark Tehranipoor.
\newblock On malicious implants in pcbs throughout the supply chain.
\newblock {\em Integration}, 2021.

\bibitem{he2017hardware}
Jiaji He, Yiqiang Zhao, Xiaolong Guo, and Yier Jin.
\newblock Hardware trojan detection through chip-free electromagnetic
  side-channel statistical analysis.
\newblock {\em IEEE Transactions on Very Large Scale Integration (VLSI)
  Systems}, 25(10):2939--2948, 2017.

\bibitem{hennessy2016jtag}
Andrew Hennessy, Yu~Zheng, and Swarup Bhunia.
\newblock Jtag-based robust pcb authentication for protection against
  counterfeiting attacks.
\newblock In {\em 2016 21st Asia and South Pacific Design Automation Conference
  (ASP-DAC)}, pages 56--61. IEEE, 2016.

\bibitem{ma4e1340_sot23}
MACOM Technology~Solutions Inc.
\newblock Ma4e1340 series datasheet.
\newblock \url{https://cdn.macom.com/datasheets/MA4E1340\%20Series.pdf}, 2021.
\newblock Accessed July 30, 2021.

\bibitem{intelpdn}
Intel.
\newblock Using the altera pdn tool to optimize your power delivery network
  design.
\newblock
  \url{https://www.intel.com/content/dam/www/programmable/us/en/pdfs/literature/an/an750.pdf},
  2015.
\newblock Accessed April 30, 2020.

\bibitem{iqbal2017pcb}
Taswar Iqbal and Kai-Dietrich Wolf.
\newblock Pcb surface fingerprints based counterfeit detection of electronic
  devices.
\newblock {\em Electronic Imaging}, 2017(7):144--149, 2017.

\bibitem{Jekel2019}
Charles~F Jekel, Gerhard Venter, Martin~P Venter, Nielen Stander, and Raphael~T
  Haftka.
\newblock {Similarity measures for identifying material parameters from
  hysteresis loops using inverse analysis}.
\newblock {\em International Journal of Material Forming}, may 2019.

\bibitem{kim2010chip}
Jaemin Kim, Woojin Lee, Yujeong Shim, Jongjoo Shim, Kiyeong Kim, Jun~So Pak,
  and Joungho Kim.
\newblock Chip-package hierarchical power distribution network modeling and
  analysis based on a segmentation method.
\newblock {\em IEEE Transactions on advanced packaging}, 33(3):647--659, 2010.

\bibitem{kim_system_2008}
Wonyoung Kim, Meeta~S. Gupta, Gu-Yeon Wei, and David Brooks.
\newblock System level analysis of fast, per-core {DVFS} using on-chip
  switching regulators.
\newblock In {\em 2008 {IEEE} 14th {International} {Symposium} on {High}
  {Performance} {Computer} {Architecture}}, pages 123--134, February 2008.
\newblock ISSN: 2378-203X.

\bibitem{mcguire2019pcb}
Matthew McGuire, Umit Ogras, and Sule Ozev.
\newblock Pcb hardware trojans: Attack modes and detection strategies.
\newblock In {\em 37th VLSI Test Symposium (VTS)}. IEEE, 2019.

\bibitem{mehta2020big}
Dhwani Mehta, Hangwei Lu, Olivia~P Paradis, Mukhil~Azhagan MS, M~Tanjidur
  Rahman, Yousef Iskander, Praveen Chawla, Damon~L Woodard, Mark Tehranipoor,
  and Navid Asadizanjani.
\newblock The big hack explained: Detection and prevention of pcb supply chain
  implants.
\newblock {\em ACM Journal on Emerging Technologies in Computing Systems
  (JETC)}, 16(4):1--25, 2020.

\bibitem{mosavirik2022scatterverif}
Tahoura Mosavirik, Fatemeh Ganji, Patrick Schaumont, and Shahin Tajik.
\newblock Scatterverif: Verification of electronic boards using reflection
  response of power distribution network.
\newblock {\em ACM Journal on Emerging Technologies in Computing Systems
  (JETC)}, 18(4):1--24, 2022.

\bibitem{mosavirik2022impedanceverif}
Tahoura Mosavirik, Patrick Schaumont, and Shahin Tajik.
\newblock Impedanceverif: On-chip impedance sensing for system-level tampering
  detection.
\newblock {\em Cryptology ePrint Archive}, 2022.

\bibitem{nishizawa2018capacitance}
Makoto Nishizawa, Kento Hasegawa, and Nozomu Togawa.
\newblock Capacitance measurement of running hardware devices and its
  application to malicious modification detection.
\newblock In {\em 2018 IEEE Asia Pacific Conference on Circuits and Systems
  (APCCAS)}, pages 362--365. IEEE, 2018.

\bibitem{novak2007frequency}
Dr~Istvan Novak and Jason~R Miller.
\newblock {\em Frequency-Domain Characterization of Power Distribution
  Networks}, volume~1.
\newblock Artech house, 2007.

\bibitem{dod2020budget}
Office of~the Under Secretary~of Defense (Comptroller)/~CFO.
\newblock Dod budget request, 2020.

\bibitem{paley2016active}
Steven Paley, Tamzidul Hoque, and Swarup Bhunia.
\newblock Active protection against pcb physical tampering.
\newblock In {\em 2016 17th International Symposium on Quality Electronic
  Design (ISQED)}, pages 356--361. IEEE, 2016.

\bibitem{paul2021silverin}
Shubhra~Deb Paul and Swarup Bhunia.
\newblock Silverin: Systematic integrity verification of printed circuit board
  using jtag infrastructure.
\newblock {\em ACM Journal on Emerging Technologies in Computing Systems
  (JETC)}, 17(3):1--28, 2021.

\bibitem{pecht2006bogus}
Michael Pecht and Sanjay Tiku.
\newblock Bogus: electronic manufacturing and consumers confront a rising tide
  of counterfeit electronics.
\newblock {\em IEEE spectrum}, 43(5):37--46, 2006.

\bibitem{piliposyan2020hardware}
Gor Piliposyan, Saqib Khursheed, and Daniele Rossi.
\newblock Hardware trojan detection on a pcb through differential power
  monitoring.
\newblock {\em IEEE Transactions on Emerging Topics in Computing}, 2020.

\bibitem{ren_frequency-dependent_2009}
Liehui Ren, Jingook Kim, Gang Feng, Bruce Archambeault, James~L. Knighten,
  James Drewniak, and Jun Fan.
\newblock Frequency-dependent via inductances for accurate power distribution
  network modeling.
\newblock In {\em 2009 {IEEE} {International} {Symposium} on {Electromagnetic}
  {Compatibility}}, pages 63--68, August 2009.
\newblock ISSN: 2158-1118.

\bibitem{robertson2018big}
Jordan Robertson and Michael Riley.
\newblock The big hack: How china used a tiny chip to infiltrate us companies.
\newblock {\em Bloomberg Businessweek}, 4, 2018.

\bibitem{samsung2019supply}
Samsung.
\newblock A journey towards a sustainable future: Sustainability in the samsung
  supply chain, 2019.

\bibitem{sandler2016extending}
Steven~M Sandler.
\newblock Extending the usable range of the 2-port shunt through impedance
  measurement.
\newblock In {\em 2016 IEEE MTT-S Latin America Microwave Conference (LAMC)},
  pages 1--3. IEEE, 2016.

\bibitem{shwartz2020inner}
Omer Shwartz, Amir Cohen, Asaf Shabtai, and Yossi Oren.
\newblock Inner conflict: How smart device components can cause harm.
\newblock {\em Computers \& Security}, 89:101665, 2020.

\bibitem{smith_2021}
Adam Smith.
\newblock H.r.6395 - 116th congress (2019-2020): National defense authorization
  act for fiscal year 2021, May 2021.

\bibitem{soman2020development}
Rohan Soman, Shishir~Kumar Singh, Tomasz Wandowski, and Pawel Malinowski.
\newblock Development of robust metric based on cumulative electrical power for
  electromechanical impedance based structural health monitoring.
\newblock {\em Smart Materials and Structures}, 29(11):115047, 2020.

\bibitem{staat2022anti}
Paul Staat, Johannes Tobisch, Christian Zenger, and Christof Paar.
\newblock Anti-tamper radio: System-level tamper detection for computing
  systems.
\newblock In {\em 2022 IEEE Symposium on Security and Privacy (SP)}, pages
  1722--1736. IEEE, 2022.

\bibitem{comparison_intel_i350}
STH.
\newblock Comparison: Intel i350-t4 genuine vs fake.
\newblock
  \url{https://forums.servethehome.com/index.php?threads/comparison-intel-i350-t4-genuine-vs-fake.6917/},
  2015.
\newblock Accessed April, 2021.

\bibitem{suh2003efficient}
G~Edward Suh, Dwaine Clarke, Blaise Gasend, Marten Van~Dijk, and Srinivas
  Devadas.
\newblock Efficient memory integrity verification and encryption for secure
  processors.
\newblock In {\em Proceedings. 36th Annual IEEE/ACM International Symposium on
  Microarchitecture, 2003. MICRO-36.}, pages 339--350. IEEE, 2003.

\bibitem{swaminathan_power_2004}
M.~Swaminathan, Joungho Kim, I.~Novak, and J.P. Libous.
\newblock Power distribution networks for system-on-package: status and
  challenges.
\newblock {\em IEEE Transactions on Advanced Packaging}, 27(2):286--300, May
  2004.
\newblock Conference Name: IEEE Transactions on Advanced Packaging.

\bibitem{mark2017honda}
Mark Tehranipoor and et.al.
\newblock Invasion of the hardware snatchers cloned electronics pollute the
  market.
\newblock
  \url{https://spectrum.ieee.org/invasion-of-the-hardware-snatchers-cloned-electronics-pollute-the-market},
  2017.

\bibitem{wang2019system}
Xiaoxiao Wang, Yueying Han, and Mark Tehranipoor.
\newblock System-level counterfeit detection using on-chip ring oscillator
  array.
\newblock {\em IEEE Transactions on Very Large Scale Integration (VLSI)
  Systems}, 27(12):2884--2896, 2019.

\bibitem{wei2018know}
Lingxiao Wei, Bo~Luo, Yu~Li, Yannan Liu, and Qiang Xu.
\newblock I know what you see: Power side-channel attack on convolutional
  neural network accelerators.
\newblock In {\em Proceedings of the 34th Annual Computer Security Applications
  Conference}, pages 393--406, 2018.

\bibitem{wei2019transmission}
Tao Wei and Jie Huang.
\newblock Transmission line identification via impedance inhomogeneity pattern.
\newblock {\em IEEE Journal of Radio Frequency Identification}, 3(4):245--251,
  2019.

\bibitem{battery_size}
Wikipedia.
\newblock List of battery sizes, November 2021.

\bibitem{xie2013detecting}
Feng Xie, Alexandra Uitdenbogerd, and Andy Song.
\newblock Detecting pcb component placement defects by genetic programming.
\newblock In {\em 2013 IEEE Congress on Evolutionary Computation}, pages
  1138--1145. IEEE, 2013.

\bibitem{xilinxdeviceguide}
Xilinx.
\newblock Device package user guide ug112.
\newblock
  \url{https://www.xilinx.com/support/documentation/user_guides/ug112.pdf},
  2012.
\newblock Accessed April, 2021.

\bibitem{xu2020bus}
Zhenyu Xu, Thomas Mauldin, Zheyi Yao, Shuyi Pei, Tao Wei, and Qing Yang.
\newblock A bus authentication and anti-probing architecture extending hardware
  trusted computing base off cpu chips and beyond.
\newblock In {\em 2020 ACM/IEEE 47th Annual International Symposium on Computer
  Architecture (ISCA)}, pages 749--761. IEEE, 2020.

\bibitem{zhang2015robust}
Fengchao Zhang, Andrew Hennessy, and Swarup Bhunia.
\newblock Robust counterfeit pcb detection exploiting intrinsic trace impedance
  variations.
\newblock In {\em 2015 IEEE 33rd VLSI Test Symposium (VTS)}, pages 1--6. IEEE,
  2015.

\bibitem{zhu2020powerscout}
Huifeng Zhu, Xiaolong Guo, Yier Jin, and Xuan Zhang.
\newblock Powerscout: A security-oriented power delivery network modeling
  framework for cross-domain side-channel analysis.
\newblock In {\em 2020 Asian Hardware Oriented Security and Trust Symposium
  (AsianHOST)}, pages 1--6. IEEE, 2020.

\bibitem{zhu2021pcbench}
Huifeng Zhu, Xiaolong Guo, Yier Jin, and Xuan Zhang.
\newblock Pcbench: Benchmarking of board-level hardware attacks and trojans.
\newblock In {\em 2021 26th Asia and South Pacific Design Automation Conference
  (ASP-DAC)}, pages 396--401. IEEE, 2021.

\end{thebibliography}
